\documentclass[superscriptaddress,reprint,prb]{revtex4-1}
\usepackage{graphicx}
\usepackage[margin=0.7in]{geometry}
\usepackage{amsmath}
\usepackage{hyperref}
\usepackage{multirow}

\let\hatOrig\hat
\renewcommand{\vec}[1]{\boldsymbol{\mathbf{#1}}}
\renewcommand{\hat}[1]{\boldsymbol{\mathbf{\hatOrig{#1}}}}

\newcommand{\sub}[1]{\ensuremath{_{\textrm{#1}}}} \newcommand{\super}[1]{\ensuremath{^{\textrm{#1}}}}  

\newcommand{\RPIMSE}{Department of Materials Science and Engineering, Rensselaer Polytechnic Institute, 110 8\super{th} Street, Troy, NY}

\begin{document}

\title{Substrate effects on charged defects in two-dimensional materials}

\author{Dan Wang}\affiliation{\RPIMSE}
\author{Ravishankar Sundararaman}\email{sundar@rpi.edu}\affiliation{\RPIMSE}

\date{\today}

\begin{abstract}
Two-dimensional (2D) materials are strongly affected by the dielectric environment including substrates,
making it an important factor in designing materials for quantum and electronic technologies.
Yet, first-principles evaluation of charged defect energetics in 2D materials typically do not
include substrates due to the high computational cost. 
We present a general continuum model approach to incorporate substrate effects directly
in density-functional theory calculations of charged defects in the 2D material alone.
We show that this technique accurately predicts charge defect energies compared to
much more expensive explicit substrate calculations, but with the computational
expediency of calculating defects in free-standing 2D materials.
Using this technique, we rapidly predict the substantial modification of charge transition levels
of two defects in MoS\sub{2} and ten defects promising for quantum technologies in hBN,
due to SiO\sub{2} and diamond substrates.
This establishes a foundation for high-throughput computational screening of 
new quantum defects in 2D materials that critically accounts for substrate effects.
\end{abstract}
\maketitle

\section{Introduction}
Point defects such as vacancies and substitutional impurities play a central role in determining
the opto-electronic properties of 2D materials desirable for electronic devices and quantum information
applications.\cite{toth2019single,lin2016defect, wang_engineering_2017, rasool2015atomic, hong2017atomic,alkauskas2016tutorial}
Their versatile functionality ranges from providing free carriers for charge transport in 2D
semiconductors\cite{zhang2002microscopic,naik_substrate_2018,singh_engineering_2018}
to encoding information in spin states for compact solid-state qubits.\cite{tawfik_first-principles_2017,
sajid_defect_2018,gupta_two-level_2019,grosso_tunable_2017,tran_robust_2016} 
The complexity of controllably synthesizing, identifying and measuring properties of point defects
necessitates first-principles computational predictions based on density-functional theory (DFT)
to first screen for desirable defects and predict experimental signatures to aid their identification.
In 2D materials, calculating energies of charged defects is complicated by the weak
and highly anisotropic screening in these systems.\cite{komsa_finite-size_2013}
The energy of a 2D supercell containing a charged defect diverges with cell size due to
strong Coulomb interactions of the defect charge with its periodic images
and compensating background.\cite{wang_determination_2015}
Several complementary approaches specialized for charged defects in 2D
materials\cite{komsa_finite-size_2013,wang_determination_2015,sundararaman_first-principles_2017}
have made it possible to reliably predict charge transition levels and engineer
defects in free-standing 2D materials.\cite{wu_first-principles_2017,wang_charged_2017,
freysoldt_first-principles_2018,wang_native_2017,naik2018coffee,smart_fundamental_2018,wang_excitation_2019}
 
However, 2D materials in most experiments and device configurations are not
free-standing and are instead deposited, grown or transferred onto a substrate.
The substrate is typically an integral part of developing and utilizing 2D materials,
critical for nucleation during synthesis and mechanical stability in operation,
and it is an unavoidable modification introduced to tune its properties.
A thorough understanding of defect properties in 2D materials would therefore
be unattainable without taking substrate effects into account.
Yet, most computational studies of defects in 2D materials to date
ignore substrates primarily due to the extremely high computational cost.
Taking the example of monolayer MoS\sub{2} on an SiO\sub{2} substrate here,
a typical 6$\times$6 defect supercell calculation of a free-standing monolayer would require 108 atoms,
but including a substrate with a minimal slab of SiO\sub{2}(0001) with six atomic layers increases
this to 428 atoms with a 60$\times$ increase in computational cost of plane-wave DFT calculations.
Repeating such calculations for a large number of substitutional or interstitial impurities
with different elements, vacancy configurations and complexes there-of in order to identify
ideal defect candidates \emph{on specific substrates} remains a formidable challenge.

One approach to eliminate this problem would be to remove the substrate atoms from the
DFT calculations, and instead approximately capture their effect on the 2D material and defect.
Electronic structure calculations in liquid and electrochemical environments have long had to
deal with large numbers of \emph{environment} atoms: practical approaches replace the liquid
environment with the response of an appropriately determined dielectric cavity.\cite{PCM-Review,PCM-SMD,PCM-SCCS}
Recent developments of such techniques have facilitated accurate first-principles calculations of complex
chemical processes in electrochemical environments, with virtually insignificant computational expense
beyond conventional DFT calculations in vacuum.\cite{Goddard,HeadGordon,
sundararaman_improving_2018,sundararaman_weighted-density_2014,gunceler_importance_2013,
sundararaman_spicing_2015,sundararaman_charge-asymmetric_2015}
Analogously, continuum models of substrates could enable rapid computational design
of defects in realistic 2D material configurations that include substrates.

In this paper, we present a continuum model approach for capturing substrate effects
in DFT calculations of the 2D material alone, which combined with charged defect correction schemes,
provides an efficient and general method for evaluating charged defects in realistic 2D material configurations.
We benchmark this methodology by predicting ionization energies of Re and Nb substitution defects (Re\sub{Mo} and Nb\sub{Mo})
in MoS\sub{2} on substrate SiO\sub{2} (MoS\sub{2}/SiO\sub{2}) and find the lowering of ionization energy due to increased
screening from the substrate to be in excellent agreement with DFT calculations that explicitly include the substrate.
We then use the so-proven method to predict the transition levels of ten promising defects in hBN
on SiO\sub{2} and diamond substrates (hBN/SiO\sub{2} and hBN/Diamond), and show that 
these defect levels remain deep enough for applications in quantum technologies.

\section{Theory and Methods}

\subsection{Charge transition level and Ionization Energy}
The formation energy of a defect with charge $q$ is\cite{zhang_chemical_1991,han_deep_2013}
\begin{equation}
E_{f}(q) = E\sub{tot}(q) - E\sub{host} + \sum_{i} N_{i} \mu _{i} + q \mu _{e},
\label{eqn:Ef}
\end{equation} 
where $E\sub{tot}(q)$ and $E_{host}$ are the total energies of the 
material with and without the defect, involving an exchange of $N_i$
atoms of each species $i$ with chemical potential $\mu_i$.
The electron chemical potential $\mu_e$ ranges from the valence band maximum
$\varepsilon\sub{VBM}$ to the conduction band minimum $\varepsilon\sub{CBM}$.

The calculation of $E\sub{tot}(q)$ involves a supercell with a net charge, which requires
a scheme for correcting the diverging Coulomb interaction energy with periodic images.
We employ the model-charge correction scheme described in detail in Refs.~\citenum{sundararaman_first-principles_2017} and \citenum{wu_first-principles_2017}.
Briefly, this technique corrects the energy and potential of the periodic DFT calculation
by comparing Poisson equation solutions for a spherical Gaussian model of the
defect charge interacting with a planar model for the anisotropic dielectric response
of the material in periodic versus isolated boundary conditions. 
The anisotropic dielectric function of the 2D material is also calculated
from first principles as described in Ref.~\citenum{wu_first-principles_2017}.
(See Supplemental Material for details.)
We previously showed this technique to be the most robust for 2D materials,
requiring no empirical parameters or cell-size extrapolation, and with all quantities
extracted purely from DFT calculations of the material.\cite{wu_first-principles_2017}

Once we can calculate individual charged defect formation energies using this scheme,
we can evaluate the charge transition level (CTL) of the defect, defined as the
electron chemical potential $\mu_e$ at which two adjacent charge states $q$ and $q'$
have equal formation energy.
Solving for $\mu_e$ from $E_f(q) = E_f(q')$ using (\ref{eqn:Ef}) yields
\begin{equation}
\mu(q|q') = \frac{E\sub{tot}(q) - E\sub{tot}(q')}{q'-q}.
\end{equation}
For donor defects which transition from $q=+1$ to $q=0$,
the transition level relative to the CBM is the donor ionization energy,
\begin{equation}
\textrm{IE}_d \equiv \varepsilon\sub{CBM} - \mu(+1|0) = E\sub{tot}(+1) - E\sub{tot}(0) + \varepsilon\sub{CBM},
\label{eqn:IEdonor}
\end{equation}
while for acceptor defects which transition from $q=0$ to $q=-1$,
the transition level relative to the VBM is the acceptor ionization energy,
\begin{equation}
\textrm{IE}_a \equiv \mu(0|-1) - \varepsilon\sub{VBM} = E\sub{tot}(-1) - E\sub{tot}(0) - \varepsilon\sub{VBM}.
\label{eqn:IEacceptor}
\end{equation}
Substrates strongly influence these ionization energies of defects in 2D materials,
and we evaluate these in selected test cases by directly computing $E\sub{tot}$
of a system containing the 2D material, defect and the substrate.
However, such calculations are extremely expensive and we need a technique
to account for substrate effects at reduced computational expense.
 
\subsection{Continuum model for substrate effects}

The challenge of accounting for a large number of atoms in an environment,
analogous to the substrate in the present case, has been addressed extensively
using continuum methods for capturing solvent effects in liquid-phase
electronic structure calculations.\cite{PCM-Review,PCM-SMD,PCM-SCCS,
gunceler_importance_2013,sundararaman_spicing_2015,sundararaman_charge-asymmetric_2015}
While these continuum solvation techniques vary greatly in details, 
they share one common aspect: they capture the dominant electrostatic interaction 
of the environment by placing the `solute' system in a dielectric cavity.
The dielectric bound charge induced at the surface of this cavity then
approximates the induced charges in the environment atoms, which are
now removed from the electronic structure calculation.
These models parametrize the cavity, often described in terms of a
smooth cavity shape function $s(\vec{r})$ that goes from 0 in the solute region
to 1 in the solvent (environment) region,\cite{gunceler_importance_2013}
and constrain parameters by fitting to solvation free energies determined
from temperature-dependent solubility measurements.

\begin{figure}
\includegraphics[width=\columnwidth]{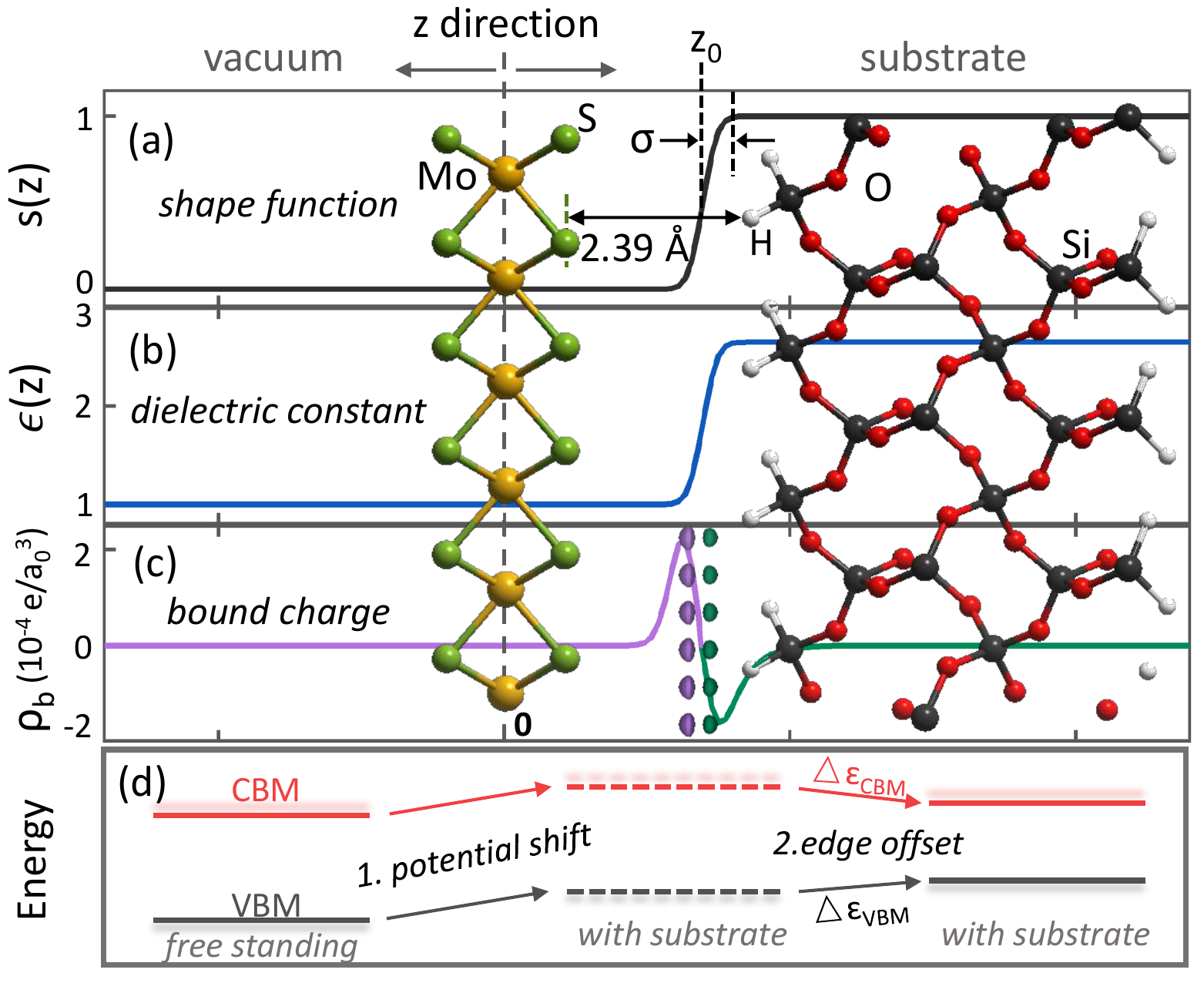}
\caption{Continuum model of SiO\sub{2} substrate specified by
(a) `shape function' $s(z)$ switching from 0 outside substrate to 1 within,
transitioning with width $\sigma$ centered at location $z_0$,
relative to MoS\sub{2} monolayer centered at $z=0$.
(b) Corresponding dielectric function and (c) bound charge induced in substrate.
(d) Scheme for evaluating substrate effects on band edge positions,
accounting for rigid shift due to the substrate potential from the continuum calculation
and edge shifts ($\Delta\varepsilon\sub{CBM}$, $\Delta\varepsilon\sub{VBM}$)
from an explicit calculation of the perfect 2D material on the substrate.
\label{fig:Schematic}}
\end{figure}

We can similarly replace the dielectric effect of the substrate by replacing it
with a dielectric slab described by a smooth shape function (Fig.~\ref{fig:Schematic}(a)),
\begin{equation}
s(z) =\frac{1}{2}\mathrm{erfc}\left(\frac{z_{0}-z}{\sigma\sqrt{2}}\right),
\end{equation}
which modulates the environment dielectric constant (Fig.~\ref{fig:Schematic}(b))
as $\epsilon(z) = 1 + (\epsilon_b - 1) s(z)$, where $\epsilon_b$ is the
bulk dielectric constant of the substrate. 
Note that the appropriate value of the bulk dielectric constant is the
optical dielectric constant ($\epsilon_\infty$) if the substrate atoms are not
allowed to relax, and the low-frequency value if atomic relaxations are allowed.
Here, we used the optical value for all cases, because we do not relax substrate
atomic geometry for each defect configuration for computational expediency.
In the example of MoS\sub{2} on SiO\sub{2} shown in Fig.~\ref{fig:Schematic},
the MoS\sub{2} monolayer is centered at $z=0$, the substrate dielectric function
smoothly `turns on' centered at $z=z_0$ over a width controlled by $\sigma$,
to the bulk value of $\epsilon_b = 2.65$ deep within the substrate.
(The resulting thicknesses of the vacuum and dielectric slab regions are
$L_z/2 + z_0 $ and $L_z/2 - z_0$ respectively, where $L_z$ is the length of the calculation
cell normal to the 2D material, as shown in the figure S1 of the Supplemental Material.)
Self-consistent solution of the modified Poisson equation with this dielectric profile
replaces the Hartree term in the DFT calculation,\cite{gunceler_importance_2013}
and produces the bound charge at the surface of the continuum substrate
shown in Fig.~\ref{fig:Schematic}(c).

This substrate continuum model involves as-yet undetermined parameters $\sigma$ and $z_0$
which both affect the proximity of the substrate dielectric response to the 2D material.
We then constrain the continuum model parameters to reproduce the response of
an explicit DFT substrate to charge distributions in the 2D material.
First, we calculate the interaction energy of the Gaussian test charge with the DFT substrate,
\begin{equation}
E\sub{int}\super{DFT} = E_{s+g} - E_s - E_g,
\end{equation}
where $E_{s+g}$ and $E_s$ are DFT energies of the substrate alone, with and without
an external Gaussian test charge placed at $z=0$ (the center of the 2D material),
and $E_g$ is the electrostatic self energy of the Gaussian charge alone.
Note that we use the model-charge-based correction scheme to handle the net charge
in the supercell calculation of $E_{s+g}$, exactly as for the charged defects.\cite{sundararaman_first-principles_2017}
Next, for a given dielectric profile based on the cavity shape function $s(z)$,
we can directly calculate the interaction energy $E\sub{int}^{s(z)}$ of the test charge with
the continuum dielectric by solving the modified Poisson equation in cylindrical coordinates,
which we do using a Bessel function expansion as described in detail in Ref.~\citenum{sundararaman_first-principles_2017}.
Finally, we select the cavity parameters such that $E\sub{int}^{s(z)} = E\sub{int}\super{DFT}$.
However, we have two parameters $\sigma$ and $z_0$, so we fix $\sigma$
and determine $z_0$ to satisfy the above condition.
(The resulting values of $z_0$ are listed in Table S1 of the Supplemental Material.)
Fortunately, we find that the predictions for charged defects are independent of $\sigma$ once
it is smaller than $\sim 0.3~\AA$, as detailed below in the discussion of Fig.~\ref{fig:MoS2}.
In summary, we use two DFT calculations of the substrate alone to determine the
continuum model parameters, which can then be used for systematically studying the
impact of that substrate on several charged defect configurations in 2D materials.

Beyond the electrostatic interaction of the defect captured by the substrate continuum model,
the substrate modifies the electronic band structure of the 2D material itself, which is
vital to capture because the defect ionization energies given by (\ref{eqn:IEdonor})
and (\ref{eqn:IEacceptor}) depend on the CBM and VBM energies respectively.
Fig.~\ref{fig:Schematic}(d) summarizes our approach to capture this effect.
First, the continuum model accounts for the overall electrostatic potential shift
at the location of the 2D material due to the substrate, which shifts the CBM
and VBM equally, but does not change the band gap.
Next, by aligning core levels (which are sensitive only to electrostatic potential)
in density-of-states calculations of the substrate and substrate + perfect 2D material, 
we can identify the shifts of the VBM and CBM that are \emph{beyond} electrostatic potential effects.
Putting these together, we get the VBM and CBM shifts in the 2D material due to both the
overall electrostatic potential and electronic effects beyond it.
In the specific example of MoS\sub{2}/SiO\sub{2} shown in Fig.~\ref{fig:Schematic}(d),
we find the offsets to be $\Delta\varepsilon\sub{VBM}$ = +0.056 eV and
$\Delta\varepsilon\sub{CBM}$ = -0.008 eV for a net band gap reduction of
0.064 eV due to the SiO\sub{2} substrate.
We illustrate this process in greater detail below for hBN on SiO\sub{2} and diamond,
including with projected band structures to isolate the 2D material band structure on a substrate.
Once again, our overall calculation procedure using the continuum methodology proposed here
involves only two calculations of the substrate alone and one of the substrate with a perfect 2D material.
Importantly, these calculations are required only once for the 2D material and substrate combination,
and no explicit substrates are included in the large supercell calculations \emph{per defect configuration}.

\subsection{Computational details}

We implemented the above technique and performed all calculations below
in the open-source plane-wave DFT software, JDFTx.\cite{sundararaman_jdftx:_2017}
We used the Garrity-Bennett-Rabe-Vanderbilt ultrasoft pseudopotentials
at their recommended kinetic energy cutoffs of 20 and 100 Hartrees for the
electronic wave function and charge density respectively.\cite{garrity2014pseudopotentials}
All supercell calculations below additionally employ Brillouin zone sampling
with a 2$\times$2 Monkhorst-Pack $k$-mesh, and truncated Coulomb
interactions to eliminate interactions with periodic images along the
slab normal `$z$' direction.\cite{sundararaman2013regularization}

We use a 6$\times$6 supercell with 30~\AA~and 16~\AA~lengths in the $z$ direction
for  MoS\sub{2} and hBN respectively, which is sufficient to completely
converge results with the truncated Coulomb interactions.
(For example, the ionization energy of the C\sub{B} defect in hBN/SiO\sub{2}
changes only by 20 meV when this length is increased from 16~\AA~to 30~\AA.)
We employ the local-density exchange-correlation functional for the MoS\sub{2} systems
in order to benchmark against previous explicit calculations,\cite{noh_deep--shallow_2015}
and the Perdew-Burke-Ernzerhof generalized gradient functional\cite{perdew1996generalized}
with DFT-D2 dispersion correction\cite{Dispersion-Grimme} for the hBN systems. 

While the technique developed here applies readily to any DFT functional or many-body method,
we employed semi-local exchange-correlation functionals to rapidly explore
several systems for developing and testing our new method,
including with large calculations of explicit substrates.
Hybrid DFT and many-body perturbation theory, which
typically predict the band edge positions and the gap with greater
accuracy,\cite{freysoldt2009controlling,huser2013quasiparticle}
may update the absolute values of ionization energies.
In particular, many-body perturbation theory techniques capture
more substantial changes in the band edge positions and gap due to
substrate screening that is not captured in DFT,\cite{naik_substrate_2018}
which then impacts the ionization energies referenced to the band edges.
However, semi-local DFT predicts the correct trends in the defect transition levels,
as shown previously for free-standing hBN,\cite{wu_first-principles_2017,smart_fundamental_2018}
and we restrict our focus to the DFT level for this initial test of the continuum methods.

For optimal lattice matching, the explicit-substrate MoS\sub{2}/SiO\sub{2} and hBN/SiO\sub{2}
calculations respectively used 4$\times$4 and 3$\times$3 supercells of $\alpha$-SiO\sub{2}(0001) slabs
with six Si atomic layers, while the hBN/Diamond calculations used a
6$\times$6 supercell of diamond(111) with eight C atomic layers.
In each case, the lateral lattice constants were set to the optimal values for the
2D material, resulting in a 5.16\%, 0.04\% and 0.50\% substrate strain
in the MoS\sub{2}/SiO\sub{2}, hBN/SiO\sub{2} and hBN/Diamond cases respectively.
All dangling bonds on the substrate surfaces were passivated with H atoms.
The atomic geometry of the substrate is optimized initially, and then
held fixed for the defect supercell calculations for computational efficiency,
while the atoms within the 2D material are fully relaxed in all calculations.
(Note that this is a convenient benchmark for the continuum model calculations
with the substrate dielectric constant set to $\epsilon_\infty$, as discussed above.
The continuum model can be used to predict results corresponding to full relaxation
by replacing $\epsilon_\infty$ with the low-frequency dielectric constant at no
additional computational cost.)

\section{Results and Discussion}

\subsection{Defects in MoS\sub{2} on SiO\sub{2}}

\begin{figure}
\includegraphics[width=\columnwidth]{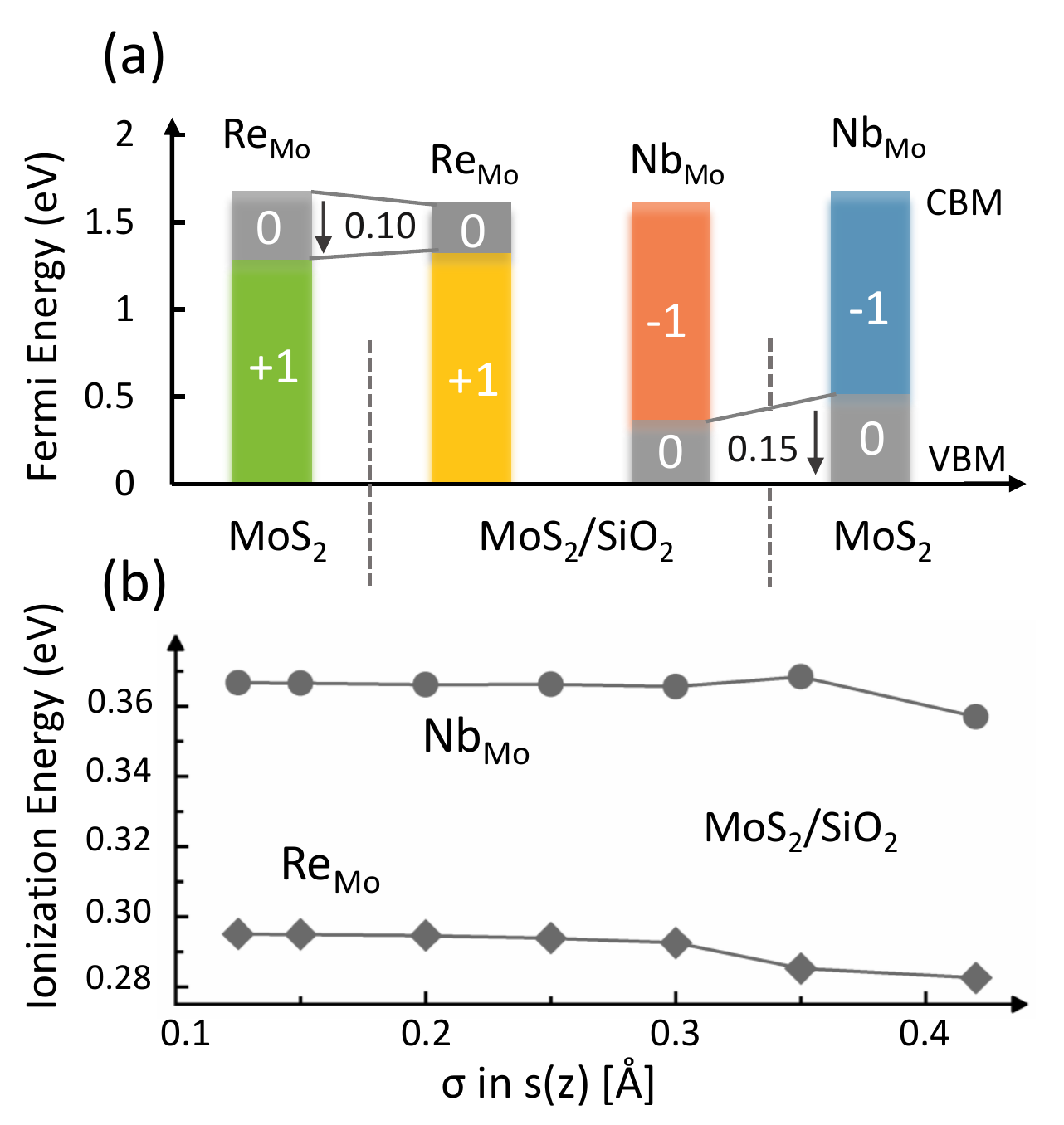}
\caption{Defect calculation in MoS\sub{2} system. (a) The stable charge state of Re\sub{Mo}
and Nb\sub{Mo} in MoS\sub{2} and MoS\sub{2}/SiO\sub{2}\ for Fermi energy ranging from VBM to CBM. 
The Fermi energy at the intersection of two different charge states ($q = +1$ and $q = 0$ for donor,
$q = 0$ and $q = -1$ for acceptor) dictates the defect transition level.
The corresponding defect ionization energies are denoted by the gray shadows.
(b) The ionization energies of Re\sub{Mo} and Nb\sub{Mo} in MoS\sub{2}/SiO\sub{2} as a function of $\sigma$.
\label{fig:MoS2}}
\end{figure}

We start by testing our technique on the only charged defects in 2D materials
for which previous calculations have explicitly included substrate effects:
Re\sub{Mo} and Nb\sub{Mo} in MoS\sub{2}/SiO\sub{2}.
Re and Nb have one more and less electron relative to Mo, so that these
substitution defects act as a donor and an acceptor respectively.
Fig.~\ref{fig:MoS2}(a) shows the prediction of the charge transition level
of these defects using the continuum methodology above.
Both defects exhibit a reduction of the defect ionization energy
by around 0.10-0.15~eV due to the SiO\sub{2} substrate, which is a significant
effect for defects that have initial ionization energies of 0.4-0.5~eV.
(See Table S2 in the Supplemental Material for a listing of calculated ionization energies.)
These continuum model results (in terms of ionization energy reduction) are in excellent agreement
of within 0.05~eV with previous results from much more expensive explicit substrate calculations,
\cite{noh_deep--shallow_2015} demonstrating the reliability of our method.
Note that here and below, we present charge transition levels and defect
ionization energies instead of the closely related charged defect formation energies
because these are easier to standardize and compare across defects; the latter also
depend on reference chemical potentials for each atom removed or added by the defect.

Fig.~\ref{fig:MoS2}(b) shows the variation of the results with the one free parameter
$\sigma$ that sets the smoothness of the transition from vacuum to substrate dielectric constant.
(As discussed above, $z_0$, which sets the center of the transition of $s(z)$, is constrained
using the response of a DFT substrate to a Gaussian test charge, for a given $\sigma$.)
The results are insensitive to $\sigma$ as long as it is small enough, with variations
in the predicted ionization energies far below 0.01~eV for $\sigma < 0.3$~\AA.
We recommend $\sigma = 0.2$~\AA~for subsequent calculations, which is small enough
to avoid overlap of 2D material charge density with the substrate dielectric response,
and yet large enough to exhibit a smooth dielectric constant variation that is
easily resolvable on a charge density grid with resolution $\sim0.1$~\AA~(kinetic
energy cutoff $\sim100$ Hartrees).

Intuitively, the substrate dielectric screening stabilizes charged states of defects,
making them easier to ionize and thereby shifting the charge transition levels
closer to the corresponding band edges (VBM for acceptor and CBM for donors).
The strong decrease of defect ionization energies in semiconductor MoS\sub{2}
is desirable for dopants for 2D electronics, as it makes it possible
to introduce charge carriers to the bands at lower temperatures.
Yet, the same effect can be undesirable for defect levels sought after
for quantum information, where distance from band edges enhances
life time of defect excited states, as we discuss next for hBN.

\subsection{Defects in hBN on SiO\sub{2} and diamond}

\begin{figure}
\includegraphics[width=\columnwidth]{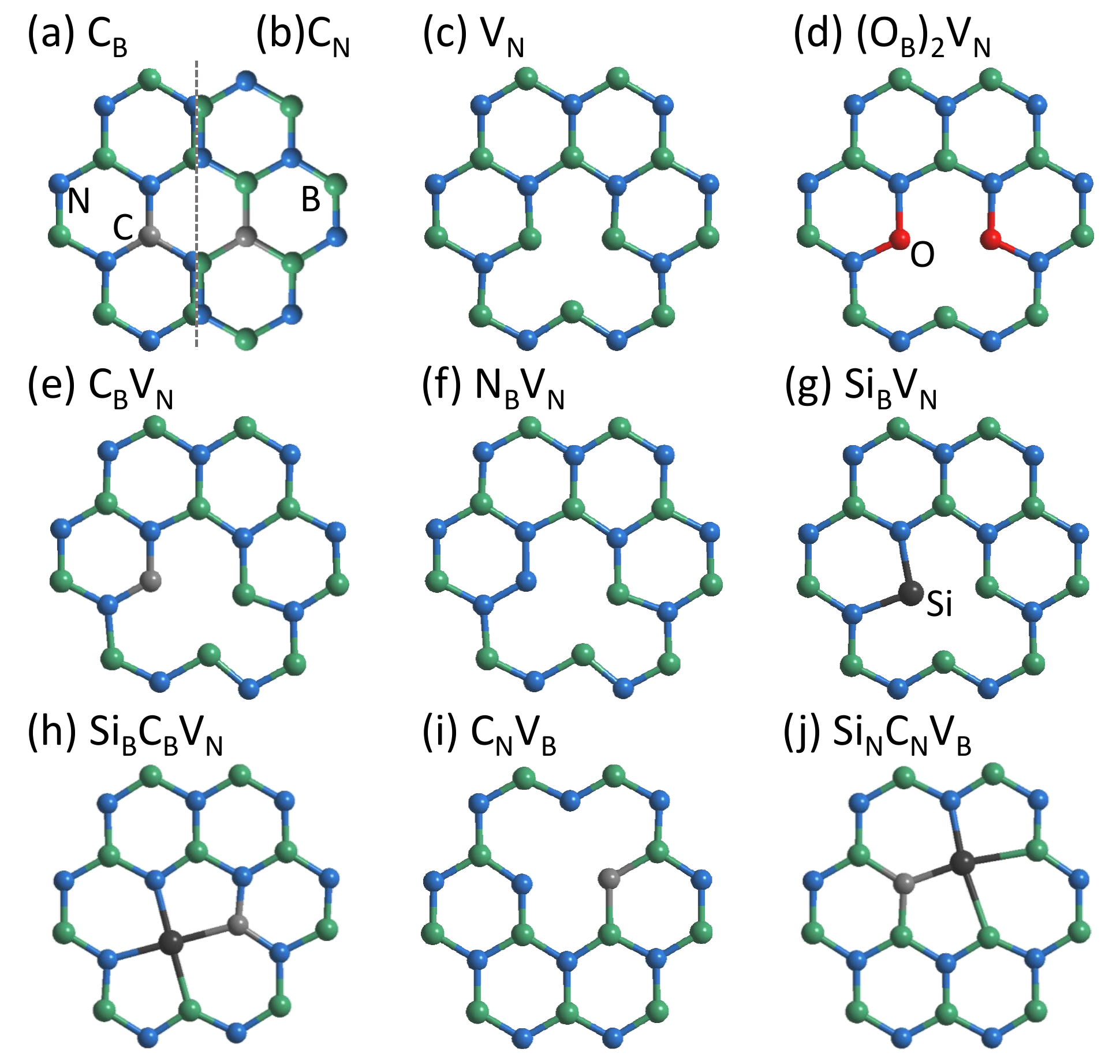}
\caption{Optimized atomic configurations of promising hBN defects considered here
using the standard notation that C\sub{B} denotes C substituting a B atom,
V\sub{N} denotes a vacancy of N, and compound defects such as C\sub{B}V\sub{N}
indicates that such substitutions / vacancies occur on adjacent atoms.
\label{fig:Configs}}
\end{figure}

Defects in hBN are the subject of increasing recent interest as candidates for single-photon
emission in a 2D analogue of the long-studied nitrogen-vacancy center in diamond.
While several recent studies have characterized the properties of spin and charge
states in hBN,\cite{wu_first-principles_2017,smart_fundamental_2018,tawfik_first-principles_2017,sajid_defect_2018,abdi2018color,reimers2018understanding}
none so far account for the effect of the substrate.
We therefore take advantage of the method established above to systematically
and rapidly investigate substrate effects on several promising hBN defects,
with atomic configurations shown in Fig.~\ref{fig:Configs}.

\begin{figure}
\includegraphics[width=\columnwidth]{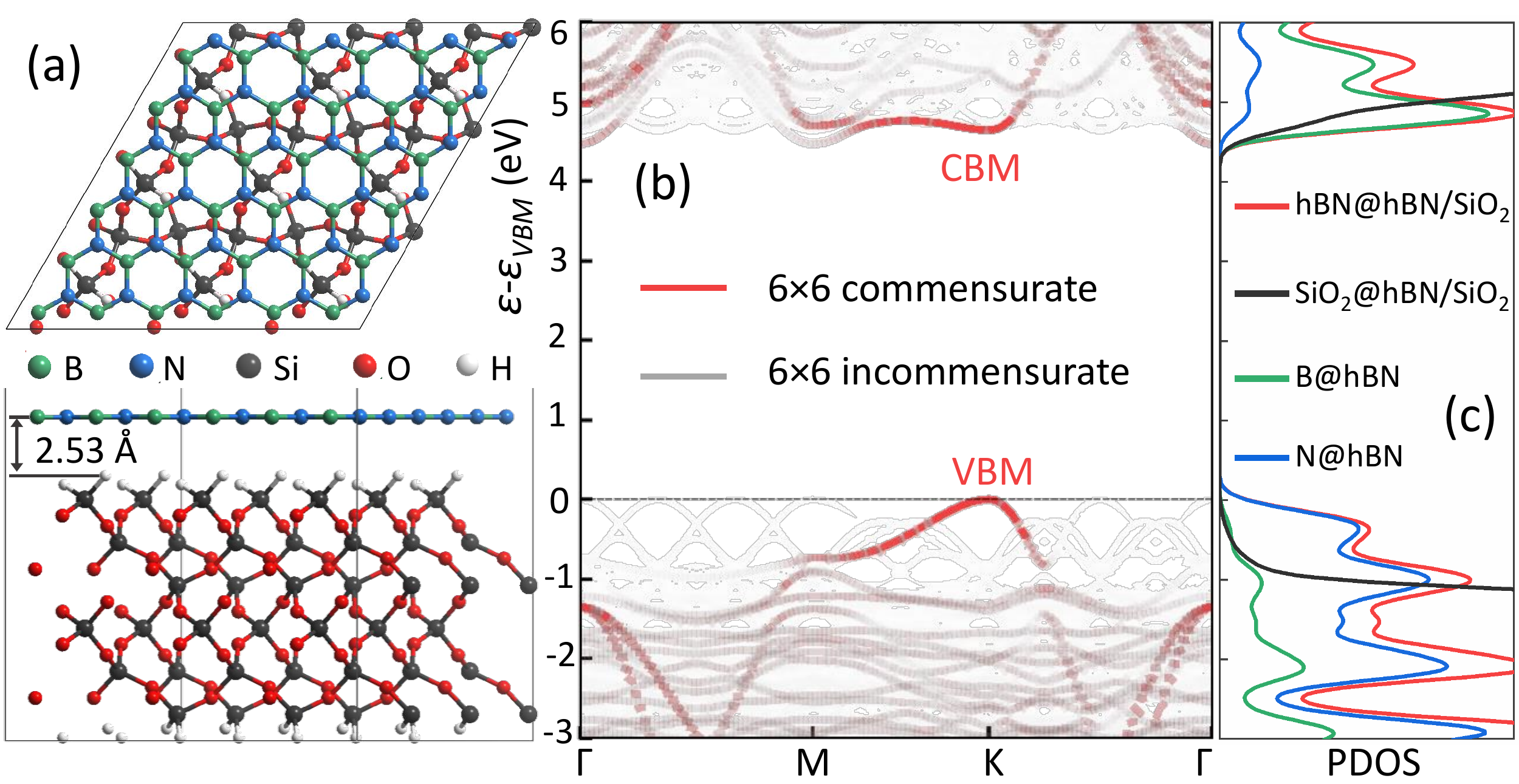}
\caption{(a) Atomic configuration of (defect-free) 6$\times$6 hBN/3$\times$3 SiO\sub{2} in top and side views.
(b) Band structure unfolded to Brillouin zone of hBN unit cell, with red and gray colors respectively indicating
components commensurate and incommensurate with the hBN unit cell, and (c) corresponding DOS.
\label{fig:hBNsilica}}
\end{figure}

\begin{figure}
\includegraphics[width=\columnwidth]{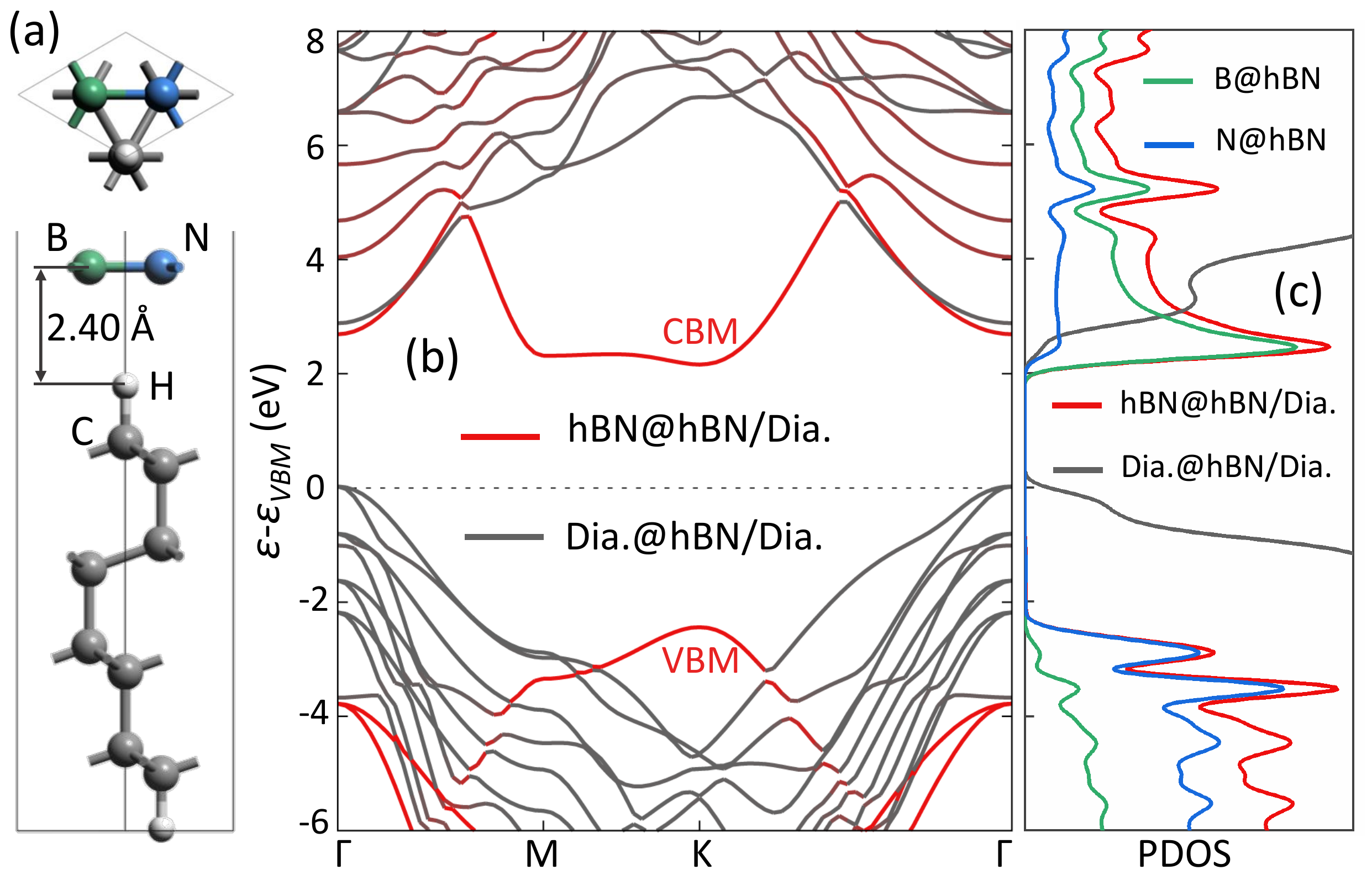}
\caption{(a) Atomic configuration of hBN/diamond in top and side views.
(b) Projected band structure with red and black colors indicating weights of hBN and
diamond atomic orbital projections respectively, and (c) corresponding DOS.
\label{fig:hBNdiamond}}
\end{figure}

First, we determine the band position changes of hBN, when placed on
SiO\sub{2}(0001) and diamond(111) substrates, which is required for
ionization energy calculations using (\ref{eqn:IEdonor}) and (\ref{eqn:IEacceptor}).
For hBN/SiO\sub{2}, we do this by calculating the density of states (DOS) and unfolding the band structure
of the 6$\times$6 hBN/3$\times$3 SiO\sub{2} supercell to the Brillouin zone of hBN unit cell.\cite{popescu2012extracting}
The unfolding clearly picks out the hBN bands that are commensurate with the unit cell,
as shown by the red lines in Fig.~\ref{fig:hBNsilica}. 
The resulting band gap is 4.64~eV, 0.04~eV smaller than in free-standing hBN.
On the other hand, hBN and diamond unit cells are already lattice matched within 0.5\%,
requiring no supercell calculations for determining band alignments.
We therefore do not require band structure unfolding in this case, and instead use orbital projections
to weight the band structure and identify hBN contributions as shown in Fig.~\ref{fig:hBNdiamond}.
Stronger dielectric screening in diamond reduces the band gap further
to 4.60 eV, as shown in Fig.~\ref{fig:hBNdiamond}(b).
 
\begin{figure}
\includegraphics[width=\columnwidth]{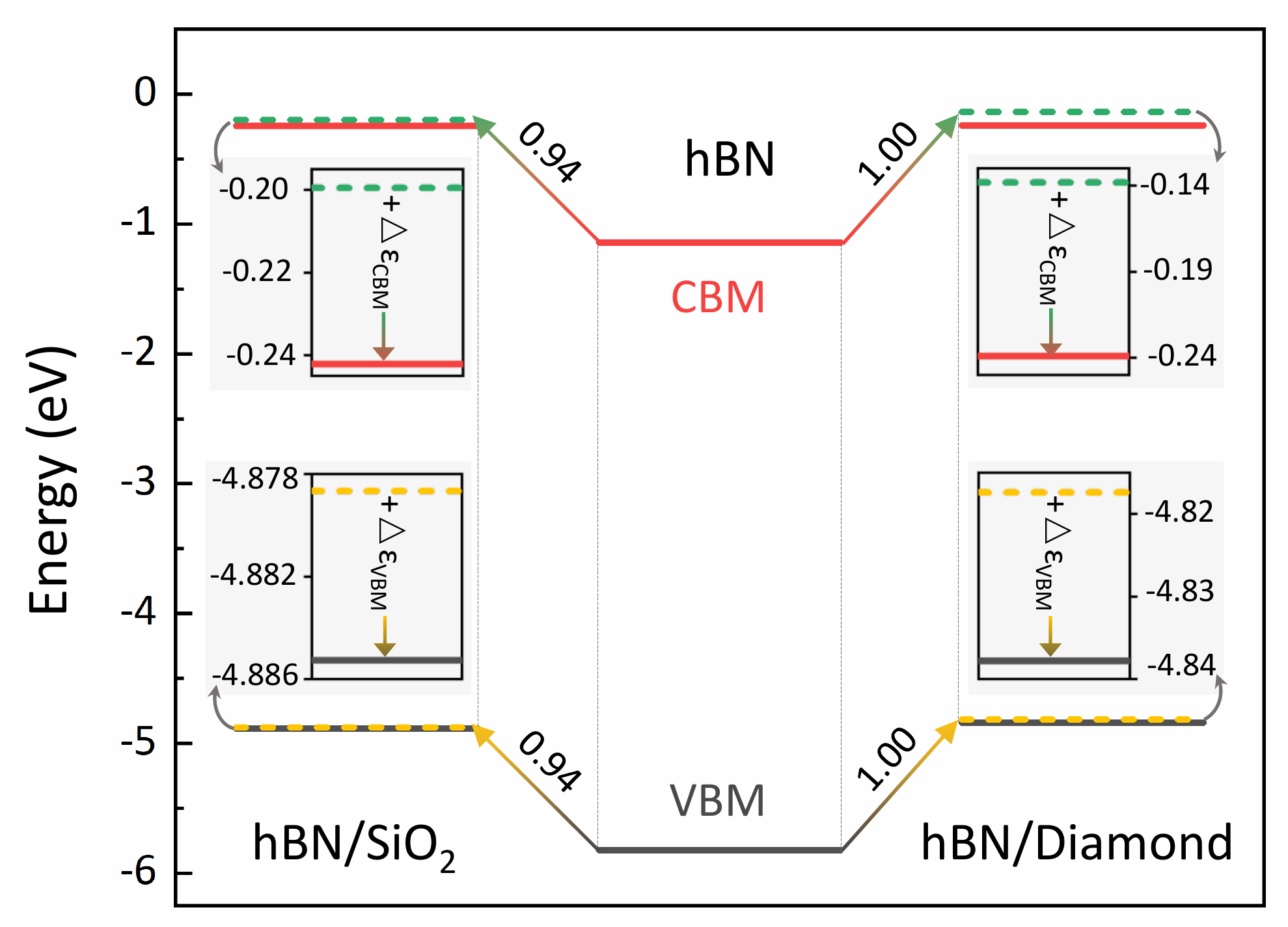}
\caption{Determination of VBM and CBM of hBN on substrates.
From hBN to hBN/SiO\sub{2}, VBM and CBM both shift up by 0.94 eV due to
the substrate electrostatic potential, and then shift $\Delta\varepsilon\sub{VBM} = -0.006$~eV
and $\Delta\varepsilon\sub{CBM} = -0.042$~eV as determined from DOS calculations
aligned by core levels (shown magnified in the insets).
Similarly, diamond introduces an electrostatic shift of 1.00~eV, followed by
$\Delta\varepsilon\sub{VBM} = -0.02$~eV and $\Delta\varepsilon\sub{CBM} = -0.10$~eV.
\label{fig:BandAlignment}}
\end{figure}

Next, after identifying the band gap modifications, we also need to determine
the band edge offsets, $\varepsilon\sub{VBM}$ and $\varepsilon\sub{CBM}$.
As discussed above, these offsets consist of an overall electrostatic potential 
shift due to the substrate that is captured by the solvation model,
specifically shifting both VBM and CBM up by 0.94 eV in hBN/SiO\sub{2}
and by 1.00 eV in hBN/Diamond relative to free-standing hBN
(dashed lines in Fig.~\ref{fig:BandAlignment}).
Further, by aligning the core levels in the DOS of isolated hBN and hBN with substrates,
we can identify the shifts in the VBM and CBM beyond the overall electrostatic potential shift.
The band edge offset is not determined from the continuum model and requires explicit DOS calibration.
This yields a $\Delta\varepsilon\sub{VBM} = -0.006$~eV and $\Delta\varepsilon\sub{CBM} = -0.042$~eV for hBN/SiO\sub{2},
and $\Delta\varepsilon\sub{VBM} = -0.02$~eV and $\Delta\varepsilon\sub{CBM} = -0.10$~eV for hBN/Diamond.
Adding these offsets yields the final reference band edges for ionization energy calculations within the continuum model
(solid lines in Fig.~\ref{fig:BandAlignment}). 

\begin{figure}
\includegraphics[width=\columnwidth]{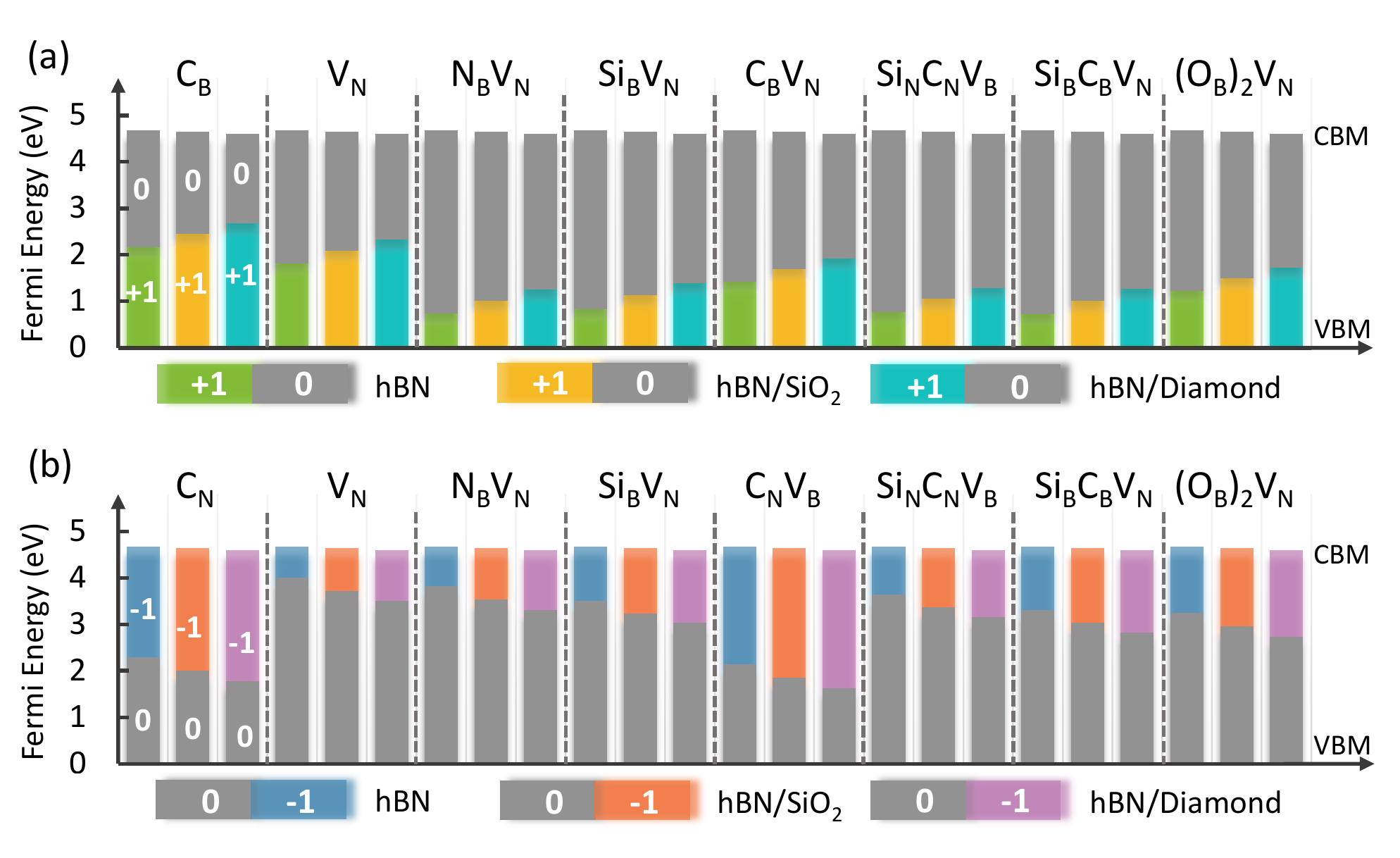}
\caption{The stable charge state of all hBN defects shown in Fig.~\ref{fig:Configs}
as a function of Fermi energy ranging from VBM to CBM.
Fermi energy at the intersection of two different charge states,
(a) $q = +1$ and $q = 0$ for donor, and (b) $q = 0$ and $q = -1$ for acceptor,
dictates the defect transition levels.
The corresponding defect ionization energies are denoted by gray shadows.
\label{fig:CTL}}
\end{figure}

The total energy calculations using the continuum model along with
the band edge positions determined above are now all we need to determine
the defect ionization energies, which are the charge
transition levels relative to the appropriate band edge.
Fig.~\ref{fig:CTL}(a) displays the calculated donor ionization energies
for several defects in hBN, hBN/SiO\sub{2} and hBN/Diamond,
while Fig.~\ref{fig:CTL}(b) shows acceptor ionization energies for several defects,
all of whose geometries are shown in Fig.~\ref{fig:Configs}.
Note that many defects are shown in both panels because they can act
as both donors and acceptors. 
The defects are all deep in hBN with ionization energies in the range of 2.14-4.01 eV.
Compared to free-standing hBN, ionization energies decrease by 0.27-0.33 eV in hBN/SiO\sub{2},
and 0.47-0.64 eV in hBN/Diamond due to increased dielectric screening by the substrate.
However, even with this systematic reduction in ionization energies, all these defect levels remain deep
-- much larger than thermal and phonon energies in the material -- indicating that they are
viable to exhibit a long coherence time even after substrate modifications to their energetics.

To confirm the accuracy of these results, we also carried out explicit 2D material + substrate calculations for a few test cases.
Specifically, we performed explicit calculations for the C\sub{B}, V\sub{N} and N\sub{B}V\sub{N}
donor ionization energies in  hBN/SiO\sub{2}, and  the donor ionization energies of C\sub{B} and V\sub{N}
as well as the acceptor ionization energy of V\sub{N} in hBN/Diamond.
We find that our continuum model predictions are accurate to within $0.10 - 0.12$~eV for the hBN/SiO\sub{2} cases,
and to within $0.11 - 0.16$~eV for hBN/diamond (see Table~S3 in the Supplemental Material for individual values).
While the absolute errors are greater than in the MoS\sub{2} case, note that the overall
magnitudes of the substrate effects are larger for hBN, resulting in a similar relative accuracy.
Similarly, formation energies of neutral defects in hBN/SiO2 and hBN/Diamond from the continuum model
are accurate to within $0.03 - 0.16$~eV of the explicit substrate calculations (see Table~S4 in the Supplemental Material),
with the difference arising mainly from effects beyond the dielectric response such as Pauli repulsion from substrate electrons.
Overall, this accuracy is remarkable considering that the continuum model calculations
required at most 72 atoms compared to 252 and 432 atoms for hBN/SiO\sub{2} and hBN/Diamond
(in addition to just 30 bohrs vacuum size compared to 56.7 and 47.2 bohrs respectively).
This amounts to a $40-200\times$ reduction in computational effort, making it now
possible to rapidly explore defect energetics with realistic treatment of substrate effects.

\subsection{Ionization energy reduction estimates}

\begin{table}
\caption{Comparison of ionization energy reduction due to substrate
predicted by self-consistent continuum model calculations for all above defects,
with a single non-self-consistent estimate for each 2D material - substrate combination
using Eq.~\ref{eqn:Estimate} with the same continuum model.
(See Table~S5 in Supplemental Material for individual results for each defect.)
\label{tab:Correction}}
\begin{tabular}{ccc}
\hline
Donors & Self-consistent & Eq.~\ref{eqn:Estimate} \\
\hline
$\Delta\rm IE\sub{MoS\sub{2}/SiO\sub{2}}$ (eV) & 0.10        & 0.10\\
$\Delta\rm IE\sub{hBN/SiO\sub{2}}$ (eV)            & $0.31-0.33$ & 0.32 \\
$\Delta\rm IE\sub{hBN/Diamond}$ (eV)               & $0.58-0.62$ & 0.60 \\
\hline
Acceptors & Continuum Model & Eq.~\ref{eqn:Estimate} \\
\hline
$\Delta\rm IE\sub{MoS\sub{2}/SiO\sub{2}}$ (eV) & 0.15        & 0.14\\
$\Delta\rm IE\sub{hBN/SiO\sub{2}}$ (eV)        & $0.27-0.29$ & 0.27 \\
$\Delta\rm IE\sub{hBN/Diamond}$ (eV)           & $0.47-0.52$ & 0.48 \\
\hline
\end{tabular}
\end{table}

We have so far presented predictions of the ionization energy of several 
donor and acceptor defects in three 2D material/substrate combinations, and compared
against explicit substrate calculations to establish the accuracy of our technique.
Table~\ref{tab:Correction} summarizes the \emph{reduction} in
ionization energy $\Delta\rm IE$ from the free-standing 2D material to
the 2D material on the substrate in each of these combinations.
Note that the reduction in ionization energy is almost the same
across all defects within each material/substrate combination.

The reason for this equivalence in ionization energy reduction is
that most charged defects have a fairly-localized charge distribution
which does not change appreciably upon the introduction of a substrate.
The charged defect corrections schemes already take advantage of this fact
to remove the periodic interaction between defects by computing the
self-energy of a Gaussian model charge in periodic and isolated
boundary conditions.\cite{komsa_finite-size_2013, sundararaman_first-principles_2017}
We could then similarly estimate the reduction in ionization energy of
charged defects as the electrostatic stabilization of a Gaussian model charge,
\begin{equation}
\Delta\textrm{IE} \approx E\super{iso}_{g}(2D) - E\super{iso}_{g}(2D/\textrm{sub}) + 
	\begin{cases}
	-\Delta\varepsilon\sub{CBM},\textrm{ Donors}\\
	+\Delta\varepsilon\sub{VBM},\textrm{ Acceptors}\\
	\end{cases}\label{eqn:Estimate}
\end{equation}
where $E\super{iso}_{g}(2D)$ and $E\super{iso}_{g}(2D/\textrm{sub})$ are the self
energies of the Gaussian model charge in isolated boundary conditions in the 
dielectric model of the 2D material alone and of the 2D material on the substrate.
(These quantities are already used in the charge defect correction for the
free-standing 2D material and 2D material on continuum model substrate respectively.)
The second term in Eq.~\ref{eqn:Estimate} accounts for the change in ionization
energy due to the substrate-induced shift of the corresponding band edge position,
which serves as the reference for defining ionization energies
(Eqs.~\ref{eqn:IEdonor} and \ref{eqn:IEacceptor}).
Note that all quantities in Eq.~\ref{eqn:Estimate} depend on the 2D material
and substrate combination alone, and not on a specific defect.

The final column of Table.~\ref{tab:Correction} shows the results of applying
Eq.~\ref{eqn:Estimate} to each of the three 2D material/substrate combinations studied above.
We find that this simple estimate agrees with a typical accuracy of 0.02~eV with
the continuum model predictions and a maximum deviation of about 0.04~eV.
(See Table~S5 in Supplemental Material additionally shows individual
results for each defect, compared with this estimate.)
This now makes it possible to rapidly estimate the ionization energy
of \emph{any} defect without even performing self-consistent DFT 
+ continuum model calculations of each defect separately.
We only need to calculate the ionization energy of all defects of interest
in a free-standing 2D material, construct the continuum model for a
substrate of interest as described above, and compute a single number using
Eq.~\ref{eqn:Estimate} to shift all free-standing defect ionization energies
to the corresponding values on substrates.

\section{Conclusions}

We have demonstrated a general framework to efficiently and accurately calculate 
energies and related properties of charged defects in 2D materials on substrates.
We resolve the challenge of first-principles evaluation of such systems by treating
the substrate as a continuous medium, with its electrostatic response 
replaced by a continuum dielectric function.
Results obtained by this method agree very well with explicit calculations that 
directly include substrate atoms, but at a small fraction of the computational effort.

This methodology applies to arbitrary combinations of defects, 2D materials and substrates,
potentially enabling high-throughput screening of not only defects with unique properties,
but also material-substrate combinations targeting desired defect properties.
As an example, application of this method to defects in MoS\sub{2} and hBN 
on SiO\sub{2} and diamond substrates reveals that enhanced screening from 
surrounding environments can significantly change transition levels.
This provides an invaluable input for the experimental identification of 2D defects for
quantum information applications such as single photon emission, fully accounting
for monolayer 2D materials on realistic substrates, as well as in
multilayer 2D materials and 2D heterostructures in future work.

\section*{Acknowledgments}
We acknowledge startup funding from the Department of Materials Science and Engineering at Rensselaer Polytechnic Institute.
All calculations were carried out at the Center for Computational Innovations at Rensselaer Polytechnic Institute.

\bibliographystyle{apsrev4-1}

\begin{thebibliography}{47}\makeatletter
\providecommand \@ifxundefined [1]{ \@ifx{#1\undefined}
}\providecommand \@ifnum [1]{ \ifnum #1\expandafter \@firstoftwo
 \else \expandafter \@secondoftwo
 \fi
}\providecommand \@ifx [1]{ \ifx #1\expandafter \@firstoftwo
 \else \expandafter \@secondoftwo
 \fi
}\providecommand \natexlab [1]{#1}\providecommand \enquote  [1]{``#1''}\providecommand \bibnamefont  [1]{#1}\providecommand \bibfnamefont [1]{#1}\providecommand \citenamefont [1]{#1}\providecommand \href@noop [0]{\@secondoftwo}\providecommand \href [0]{\begingroup \@sanitize@url \@href}\providecommand \@href[1]{\@@startlink{#1}\@@href}\providecommand \@@href[1]{\endgroup#1\@@endlink}\providecommand \@sanitize@url [0]{\catcode `\\12\catcode `\$12\catcode
  `\&12\catcode `\#12\catcode `\^12\catcode `\_12\catcode `\%12\relax}\providecommand \@@startlink[1]{}\providecommand \@@endlink[0]{}\providecommand \url  [0]{\begingroup\@sanitize@url \@url }\providecommand \@url [1]{\endgroup\@href {#1}{\urlprefix }}\providecommand \urlprefix  [0]{URL }\providecommand \Eprint [0]{\href }\providecommand \doibase [0]{http://dx.doi.org/}\providecommand \selectlanguage [0]{\@gobble}\providecommand \bibinfo  [0]{\@secondoftwo}\providecommand \bibfield  [0]{\@secondoftwo}\providecommand \translation [1]{[#1]}\providecommand \BibitemOpen [0]{}\providecommand \bibitemStop [0]{}\providecommand \bibitemNoStop [0]{.\EOS\space}\providecommand \EOS [0]{\spacefactor3000\relax}\providecommand \BibitemShut  [1]{\csname bibitem#1\endcsname}\let\auto@bib@innerbib\@empty
\bibitem [{\citenamefont {Toth}\ and\ \citenamefont
  {Aharonovich}(2019)}]{toth2019single}  \BibitemOpen
  \bibfield  {author} {\bibinfo {author} {\bibfnamefont {M.}~\bibnamefont
  {Toth}}\ and\ \bibinfo {author} {\bibfnamefont {I.}~\bibnamefont
  {Aharonovich}},\ }\href@noop {} {\bibfield  {journal} {\bibinfo  {journal}
  {Ann. Rev. Phys. Chem.}\ }\textbf {\bibinfo {volume} {70}},\ \bibinfo {pages}
  {123} (\bibinfo {year} {2019})}\BibitemShut {NoStop}\bibitem [{\citenamefont {Lin}\ \emph {et~al.}(2016)\citenamefont {Lin},
  \citenamefont {Carvalho}, \citenamefont {Kahn}, \citenamefont {Lv},
  \citenamefont {Rao}, \citenamefont {Terrones}, \citenamefont {Pimenta},\ and\
  \citenamefont {Terrones}}]{lin2016defect}  \BibitemOpen
  \bibfield  {author} {\bibinfo {author} {\bibfnamefont {Z.}~\bibnamefont
  {Lin}}, \bibinfo {author} {\bibfnamefont {B.~R.}\ \bibnamefont {Carvalho}},
  \bibinfo {author} {\bibfnamefont {E.}~\bibnamefont {Kahn}}, \bibinfo {author}
  {\bibfnamefont {R.}~\bibnamefont {Lv}}, \bibinfo {author} {\bibfnamefont
  {R.}~\bibnamefont {Rao}}, \bibinfo {author} {\bibfnamefont {H.}~\bibnamefont
  {Terrones}}, \bibinfo {author} {\bibfnamefont {M.~A.}\ \bibnamefont
  {Pimenta}}, \ and\ \bibinfo {author} {\bibfnamefont {M.}~\bibnamefont
  {Terrones}},\ }\href@noop {} {\bibfield  {journal} {\bibinfo  {journal} {2D
  Mater.}\ }\textbf {\bibinfo {volume} {3}},\ \bibinfo {pages} {022002}
  (\bibinfo {year} {2016})}\BibitemShut {NoStop}\bibitem [{\citenamefont {Wang}\ \emph
  {et~al.}(2017{\natexlab{a}})\citenamefont {Wang}, \citenamefont {Li},
  \citenamefont {Han}, \citenamefont {Tian},\ and\ \citenamefont
  {Sun}}]{wang_engineering_2017}  \BibitemOpen
  \bibfield  {author} {\bibinfo {author} {\bibfnamefont {D.}~\bibnamefont
  {Wang}}, \bibinfo {author} {\bibfnamefont {X.-B.}\ \bibnamefont {Li}},
  \bibinfo {author} {\bibfnamefont {D.}~\bibnamefont {Han}}, \bibinfo {author}
  {\bibfnamefont {W.~Q.}\ \bibnamefont {Tian}}, \ and\ \bibinfo {author}
  {\bibfnamefont {H.-B.}\ \bibnamefont {Sun}},\ }\href {\doibase
  10.1016/j.nantod.2017.07.001} {\bibfield  {journal} {\bibinfo  {journal}
  {Nano Today}\ }\textbf {\bibinfo {volume} {16}},\ \bibinfo {pages} {30}
  (\bibinfo {year} {2017}{\natexlab{a}})}\BibitemShut {NoStop}\bibitem [{\citenamefont {Rasool}\ \emph {et~al.}(2015)\citenamefont {Rasool},
  \citenamefont {Ophus},\ and\ \citenamefont {Zettl}}]{rasool2015atomic}  \BibitemOpen
  \bibfield  {author} {\bibinfo {author} {\bibfnamefont {H.~I.}\ \bibnamefont
  {Rasool}}, \bibinfo {author} {\bibfnamefont {C.}~\bibnamefont {Ophus}}, \
  and\ \bibinfo {author} {\bibfnamefont {A.}~\bibnamefont {Zettl}},\
  }\href@noop {} {\bibfield  {journal} {\bibinfo  {journal} {Adv. Mater.}\
  }\textbf {\bibinfo {volume} {27}},\ \bibinfo {pages} {5771} (\bibinfo {year}
  {2015})}\BibitemShut {NoStop}\bibitem [{\citenamefont {Hong}\ \emph {et~al.}(2017)\citenamefont {Hong},
  \citenamefont {Jin}, \citenamefont {Yuan},\ and\ \citenamefont
  {Zhang}}]{hong2017atomic}  \BibitemOpen
  \bibfield  {author} {\bibinfo {author} {\bibfnamefont {J.}~\bibnamefont
  {Hong}}, \bibinfo {author} {\bibfnamefont {C.}~\bibnamefont {Jin}}, \bibinfo
  {author} {\bibfnamefont {J.}~\bibnamefont {Yuan}}, \ and\ \bibinfo {author}
  {\bibfnamefont {Z.}~\bibnamefont {Zhang}},\ }\href@noop {} {\bibfield
  {journal} {\bibinfo  {journal} {Adv. Mater.}\ }\textbf {\bibinfo {volume}
  {29}},\ \bibinfo {pages} {1606434} (\bibinfo {year} {2017})}\BibitemShut
  {NoStop}\bibitem [{\citenamefont {Alkauskas}\ \emph {et~al.}(2016)\citenamefont
  {Alkauskas}, \citenamefont {McCluskey},\ and\ \citenamefont {Van~de
  Walle}}]{alkauskas2016tutorial}  \BibitemOpen
  \bibfield  {author} {\bibinfo {author} {\bibfnamefont {A.}~\bibnamefont
  {Alkauskas}}, \bibinfo {author} {\bibfnamefont {M.~D.}\ \bibnamefont
  {McCluskey}}, \ and\ \bibinfo {author} {\bibfnamefont {C.~G.}\ \bibnamefont
  {Van~de Walle}},\ }\href@noop {} {\bibfield  {journal} {\bibinfo  {journal}
  {J. Appl. Phys.}\ }\textbf {\bibinfo {volume} {119}},\ \bibinfo {pages}
  {181101} (\bibinfo {year} {2016})}\BibitemShut {NoStop}\bibitem [{\citenamefont {Zhang}(2002)}]{zhang2002microscopic}  \BibitemOpen
  \bibfield  {author} {\bibinfo {author} {\bibfnamefont {S.}~\bibnamefont
  {Zhang}},\ }\href@noop {} {\bibfield  {journal} {\bibinfo  {journal} {J.
  Phys.: Cond. Matt.}\ }\textbf {\bibinfo {volume} {14}},\ \bibinfo {pages}
  {R881} (\bibinfo {year} {2002})}\BibitemShut {NoStop}\bibitem [{\citenamefont {Naik}\ and\ \citenamefont
  {Jain}(2018{\natexlab{a}})}]{naik_substrate_2018}  \BibitemOpen
  \bibfield  {author} {\bibinfo {author} {\bibfnamefont {M.~H.}\ \bibnamefont
  {Naik}}\ and\ \bibinfo {author} {\bibfnamefont {M.}~\bibnamefont {Jain}},\
  }\href@noop {} {\bibfield  {journal} {\bibinfo  {journal} {Phys. Rev.
  Mater.}\ }\textbf {\bibinfo {volume} {2}},\ \bibinfo {pages} {084002}
  (\bibinfo {year} {2018}{\natexlab{a}})}\BibitemShut {NoStop}\bibitem [{\citenamefont {Singh}\ \emph {et~al.}(2018)\citenamefont {Singh},
  \citenamefont {Manjanath},\ and\ \citenamefont
  {Singh}}]{singh_engineering_2018}  \BibitemOpen
  \bibfield  {author} {\bibinfo {author} {\bibfnamefont {A.}~\bibnamefont
  {Singh}}, \bibinfo {author} {\bibfnamefont {A.}~\bibnamefont {Manjanath}}, \
  and\ \bibinfo {author} {\bibfnamefont {A.~K.}\ \bibnamefont {Singh}},\ }\href
  {\doibase 10.1021/acs.jpcc.8b08082} {\bibfield  {journal} {\bibinfo
  {journal} {J. Phys. Chem. C}\ }\textbf {\bibinfo {volume} {122}},\ \bibinfo
  {pages} {24475} (\bibinfo {year} {2018})}\BibitemShut {NoStop}\bibitem [{\citenamefont {Tawfik}\ \emph {et~al.}(2017)\citenamefont {Tawfik},
  \citenamefont {Ali}, \citenamefont {Fronzi}, \citenamefont {Kianinia},
  \citenamefont {Tran}, \citenamefont {Stampfl}, \citenamefont {Aharonovich},
  \citenamefont {Toth},\ and\ \citenamefont
  {Ford}}]{tawfik_first-principles_2017}  \BibitemOpen
  \bibfield  {author} {\bibinfo {author} {\bibfnamefont {S.~A.}\ \bibnamefont
  {Tawfik}}, \bibinfo {author} {\bibfnamefont {S.}~\bibnamefont {Ali}},
  \bibinfo {author} {\bibfnamefont {M.}~\bibnamefont {Fronzi}}, \bibinfo
  {author} {\bibfnamefont {M.}~\bibnamefont {Kianinia}}, \bibinfo {author}
  {\bibfnamefont {T.~T.}\ \bibnamefont {Tran}}, \bibinfo {author}
  {\bibfnamefont {C.}~\bibnamefont {Stampfl}}, \bibinfo {author} {\bibfnamefont
  {I.}~\bibnamefont {Aharonovich}}, \bibinfo {author} {\bibfnamefont
  {M.}~\bibnamefont {Toth}}, \ and\ \bibinfo {author} {\bibfnamefont {M.~J.}\
  \bibnamefont {Ford}},\ }\href {\doibase 10.1039/C7NR04270A} {\bibfield
  {journal} {\bibinfo  {journal} {Nanoscale}\ }\textbf {\bibinfo {volume}
  {9}},\ \bibinfo {pages} {13575} (\bibinfo {year} {2017})}\BibitemShut
  {NoStop}\bibitem [{\citenamefont {Sajid}\ \emph {et~al.}(2018)\citenamefont {Sajid},
  \citenamefont {Reimers},\ and\ \citenamefont {Ford}}]{sajid_defect_2018}  \BibitemOpen
  \bibfield  {author} {\bibinfo {author} {\bibfnamefont {A.}~\bibnamefont
  {Sajid}}, \bibinfo {author} {\bibfnamefont {J.~R.}\ \bibnamefont {Reimers}},
  \ and\ \bibinfo {author} {\bibfnamefont {M.~J.}\ \bibnamefont {Ford}},\
  }\href@noop {} {\bibfield  {journal} {\bibinfo  {journal} {Phys. Rev. B}\
  }\textbf {\bibinfo {volume} {97}},\ \bibinfo {pages} {064101} (\bibinfo
  {year} {2018})}\BibitemShut {NoStop}\bibitem [{\citenamefont {Gupta}\ \emph {et~al.}(2019)\citenamefont {Gupta},
  \citenamefont {Yang},\ and\ \citenamefont {Yakobson}}]{gupta_two-level_2019}  \BibitemOpen
  \bibfield  {author} {\bibinfo {author} {\bibfnamefont {S.}~\bibnamefont
  {Gupta}}, \bibinfo {author} {\bibfnamefont {J.-H.}\ \bibnamefont {Yang}}, \
  and\ \bibinfo {author} {\bibfnamefont {B.~I.}\ \bibnamefont {Yakobson}},\
  }\href {\doibase 10.1021/acs.nanolett.8b04159} {\bibfield  {journal}
  {\bibinfo  {journal} {Nano Lett.}\ }\textbf {\bibinfo {volume} {19}},\
  \bibinfo {pages} {408} (\bibinfo {year} {2019})}\BibitemShut {NoStop}\bibitem [{\citenamefont {Grosso}\ \emph {et~al.}(2017)\citenamefont {Grosso},
  \citenamefont {Moon}, \citenamefont {Lienhard}, \citenamefont {Ali},
  \citenamefont {Efetov}, \citenamefont {Furchi}, \citenamefont
  {Jarillo-Herrero}, \citenamefont {Ford}, \citenamefont {Aharonovich},\ and\
  \citenamefont {Englund}}]{grosso_tunable_2017}  \BibitemOpen
  \bibfield  {author} {\bibinfo {author} {\bibfnamefont {G.}~\bibnamefont
  {Grosso}}, \bibinfo {author} {\bibfnamefont {H.}~\bibnamefont {Moon}},
  \bibinfo {author} {\bibfnamefont {B.}~\bibnamefont {Lienhard}}, \bibinfo
  {author} {\bibfnamefont {S.}~\bibnamefont {Ali}}, \bibinfo {author}
  {\bibfnamefont {D.~K.}\ \bibnamefont {Efetov}}, \bibinfo {author}
  {\bibfnamefont {M.~M.}\ \bibnamefont {Furchi}}, \bibinfo {author}
  {\bibfnamefont {P.}~\bibnamefont {Jarillo-Herrero}}, \bibinfo {author}
  {\bibfnamefont {M.~J.}\ \bibnamefont {Ford}}, \bibinfo {author}
  {\bibfnamefont {I.}~\bibnamefont {Aharonovich}}, \ and\ \bibinfo {author}
  {\bibfnamefont {D.}~\bibnamefont {Englund}},\ }\href@noop {} {\bibfield
  {journal} {\bibinfo  {journal} {Nature Commun.}\ }\textbf {\bibinfo {volume}
  {8}},\ \bibinfo {pages} {705} (\bibinfo {year} {2017})}\BibitemShut {NoStop}\bibitem [{\citenamefont {Tran}\ \emph {et~al.}(2016)\citenamefont {Tran},
  \citenamefont {Elbadawi}, \citenamefont {Totonjian}, \citenamefont {Lobo},
  \citenamefont {Grosso}, \citenamefont {Moon}, \citenamefont {Englund},
  \citenamefont {Ford}, \citenamefont {Aharonovich},\ and\ \citenamefont
  {Toth}}]{tran_robust_2016}  \BibitemOpen
  \bibfield  {author} {\bibinfo {author} {\bibfnamefont {T.~T.}\ \bibnamefont
  {Tran}}, \bibinfo {author} {\bibfnamefont {C.}~\bibnamefont {Elbadawi}},
  \bibinfo {author} {\bibfnamefont {D.}~\bibnamefont {Totonjian}}, \bibinfo
  {author} {\bibfnamefont {C.~J.}\ \bibnamefont {Lobo}}, \bibinfo {author}
  {\bibfnamefont {G.}~\bibnamefont {Grosso}}, \bibinfo {author} {\bibfnamefont
  {H.}~\bibnamefont {Moon}}, \bibinfo {author} {\bibfnamefont {D.~R.}\
  \bibnamefont {Englund}}, \bibinfo {author} {\bibfnamefont {M.~J.}\
  \bibnamefont {Ford}}, \bibinfo {author} {\bibfnamefont {I.}~\bibnamefont
  {Aharonovich}}, \ and\ \bibinfo {author} {\bibfnamefont {M.}~\bibnamefont
  {Toth}},\ }\href {\doibase 10.1021/acsnano.6b03602} {\bibfield  {journal}
  {\bibinfo  {journal} {ACS Nano}\ }\textbf {\bibinfo {volume} {10}},\ \bibinfo
  {pages} {7331} (\bibinfo {year} {2016})}\BibitemShut {NoStop}\bibitem [{\citenamefont {Komsa}\ and\ \citenamefont
  {Pasquarello}(2013)}]{komsa_finite-size_2013}  \BibitemOpen
  \bibfield  {author} {\bibinfo {author} {\bibfnamefont {H.-P.}\ \bibnamefont
  {Komsa}}\ and\ \bibinfo {author} {\bibfnamefont {A.}~\bibnamefont
  {Pasquarello}},\ }\href@noop {} {\bibfield  {journal} {\bibinfo  {journal}
  {Phys. Rev. Lett.}\ }\textbf {\bibinfo {volume} {110}},\ \bibinfo {pages}
  {095505} (\bibinfo {year} {2013})}\BibitemShut {NoStop}\bibitem [{\citenamefont {Wang}\ \emph {et~al.}(2015)\citenamefont {Wang},
  \citenamefont {Han}, \citenamefont {Li}, \citenamefont {Xie}, \citenamefont
  {Chen}, \citenamefont {Tian}, \citenamefont {West}, \citenamefont {Sun},\
  and\ \citenamefont {Zhang}}]{wang_determination_2015}  \BibitemOpen
  \bibfield  {author} {\bibinfo {author} {\bibfnamefont {D.}~\bibnamefont
  {Wang}}, \bibinfo {author} {\bibfnamefont {D.}~\bibnamefont {Han}}, \bibinfo
  {author} {\bibfnamefont {X.-B.}\ \bibnamefont {Li}}, \bibinfo {author}
  {\bibfnamefont {S.-Y.}\ \bibnamefont {Xie}}, \bibinfo {author} {\bibfnamefont
  {N.-K.}\ \bibnamefont {Chen}}, \bibinfo {author} {\bibfnamefont {W.~Q.}\
  \bibnamefont {Tian}}, \bibinfo {author} {\bibfnamefont {D.}~\bibnamefont
  {West}}, \bibinfo {author} {\bibfnamefont {H.-B.}\ \bibnamefont {Sun}}, \
  and\ \bibinfo {author} {\bibfnamefont {S.}~\bibnamefont {Zhang}},\
  }\href@noop {} {\bibfield  {journal} {\bibinfo  {journal} {Phys. Rev. Lett.}\
  }\textbf {\bibinfo {volume} {114}},\ \bibinfo {pages} {196801} (\bibinfo
  {year} {2015})}\BibitemShut {NoStop}\bibitem [{\citenamefont {Sundararaman}\ and\ \citenamefont
  {Ping}(2017)}]{sundararaman_first-principles_2017}  \BibitemOpen
  \bibfield  {author} {\bibinfo {author} {\bibfnamefont {R.}~\bibnamefont
  {Sundararaman}}\ and\ \bibinfo {author} {\bibfnamefont {Y.}~\bibnamefont
  {Ping}},\ }\href {\doibase 10.1063/1.4978238} {\bibfield  {journal} {\bibinfo
   {journal} {J. Chem. Phys.}\ }\textbf {\bibinfo {volume} {146}},\ \bibinfo
  {pages} {104109} (\bibinfo {year} {2017})}\BibitemShut {NoStop}\bibitem [{\citenamefont {Wu}\ \emph {et~al.}(2017)\citenamefont {Wu},
  \citenamefont {Galatas}, \citenamefont {Sundararaman}, \citenamefont
  {Rocca},\ and\ \citenamefont {Ping}}]{wu_first-principles_2017}  \BibitemOpen
  \bibfield  {author} {\bibinfo {author} {\bibfnamefont {F.}~\bibnamefont
  {Wu}}, \bibinfo {author} {\bibfnamefont {A.}~\bibnamefont {Galatas}},
  \bibinfo {author} {\bibfnamefont {R.}~\bibnamefont {Sundararaman}}, \bibinfo
  {author} {\bibfnamefont {D.}~\bibnamefont {Rocca}}, \ and\ \bibinfo {author}
  {\bibfnamefont {Y.}~\bibnamefont {Ping}},\ }\href@noop {} {\bibfield
  {journal} {\bibinfo  {journal} {Phys. Rev. Mater.}\ }\textbf {\bibinfo
  {volume} {1}},\ \bibinfo {pages} {071001} (\bibinfo {year}
  {2017})}\BibitemShut {NoStop}\bibitem [{\citenamefont {Wang}\ \emph
  {et~al.}(2017{\natexlab{b}})\citenamefont {Wang}, \citenamefont {Han},
  \citenamefont {Li}, \citenamefont {Chen}, \citenamefont {West}, \citenamefont
  {Meunier}, \citenamefont {Zhang},\ and\ \citenamefont
  {Sun}}]{wang_charged_2017}  \BibitemOpen
  \bibfield  {author} {\bibinfo {author} {\bibfnamefont {D.}~\bibnamefont
  {Wang}}, \bibinfo {author} {\bibfnamefont {D.}~\bibnamefont {Han}}, \bibinfo
  {author} {\bibfnamefont {X.-B.}\ \bibnamefont {Li}}, \bibinfo {author}
  {\bibfnamefont {N.-K.}\ \bibnamefont {Chen}}, \bibinfo {author}
  {\bibfnamefont {D.}~\bibnamefont {West}}, \bibinfo {author} {\bibfnamefont
  {V.}~\bibnamefont {Meunier}}, \bibinfo {author} {\bibfnamefont
  {S.}~\bibnamefont {Zhang}}, \ and\ \bibinfo {author} {\bibfnamefont {H.-B.}\
  \bibnamefont {Sun}},\ }\href@noop {} {\bibfield  {journal} {\bibinfo
  {journal} {Phys. Rev. B}\ }\textbf {\bibinfo {volume} {96}},\ \bibinfo
  {pages} {155424} (\bibinfo {year} {2017}{\natexlab{b}})}\BibitemShut
  {NoStop}\bibitem [{\citenamefont {Freysoldt}\ and\ \citenamefont
  {Neugebauer}(2018)}]{freysoldt_first-principles_2018}  \BibitemOpen
  \bibfield  {author} {\bibinfo {author} {\bibfnamefont {C.}~\bibnamefont
  {Freysoldt}}\ and\ \bibinfo {author} {\bibfnamefont {J.}~\bibnamefont
  {Neugebauer}},\ }\href@noop {} {\bibfield  {journal} {\bibinfo  {journal}
  {Phys. Rev. B}\ }\textbf {\bibinfo {volume} {97}},\ \bibinfo {pages} {205425}
  (\bibinfo {year} {2018})}\BibitemShut {NoStop}\bibitem [{\citenamefont {Wang}\ \emph
  {et~al.}(2017{\natexlab{c}})\citenamefont {Wang}, \citenamefont {Li},\ and\
  \citenamefont {Sun}}]{wang_native_2017}  \BibitemOpen
  \bibfield  {author} {\bibinfo {author} {\bibfnamefont {D.}~\bibnamefont
  {Wang}}, \bibinfo {author} {\bibfnamefont {X.-B.}\ \bibnamefont {Li}}, \ and\
  \bibinfo {author} {\bibfnamefont {H.-B.}\ \bibnamefont {Sun}},\ }\href
  {\doibase 10.1039/C7NR03389C} {\bibfield  {journal} {\bibinfo  {journal}
  {Nanoscale}\ }\textbf {\bibinfo {volume} {9}},\ \bibinfo {pages} {11619}
  (\bibinfo {year} {2017}{\natexlab{c}})}\BibitemShut {NoStop}\bibitem [{\citenamefont {Naik}\ and\ \citenamefont
  {Jain}(2018{\natexlab{b}})}]{naik2018coffee}  \BibitemOpen
  \bibfield  {author} {\bibinfo {author} {\bibfnamefont {M.~H.}\ \bibnamefont
  {Naik}}\ and\ \bibinfo {author} {\bibfnamefont {M.}~\bibnamefont {Jain}},\
  }\href@noop {} {\bibfield  {journal} {\bibinfo  {journal} {Comput. Phys.
  Comm.}\ }\textbf {\bibinfo {volume} {226}},\ \bibinfo {pages} {114} (\bibinfo
  {year} {2018}{\natexlab{b}})}\BibitemShut {NoStop}\bibitem [{\citenamefont {Smart}\ \emph {et~al.}(2018)\citenamefont {Smart},
  \citenamefont {Wu}, \citenamefont {Govoni},\ and\ \citenamefont
  {Ping}}]{smart_fundamental_2018}  \BibitemOpen
  \bibfield  {author} {\bibinfo {author} {\bibfnamefont {T.~J.}\ \bibnamefont
  {Smart}}, \bibinfo {author} {\bibfnamefont {F.}~\bibnamefont {Wu}}, \bibinfo
  {author} {\bibfnamefont {M.}~\bibnamefont {Govoni}}, \ and\ \bibinfo {author}
  {\bibfnamefont {Y.}~\bibnamefont {Ping}},\ }\href@noop {} {\bibfield
  {journal} {\bibinfo  {journal} {Phys. Rev. Mater.}\ }\textbf {\bibinfo
  {volume} {2}},\ \bibinfo {pages} {124002} (\bibinfo {year}
  {2018})}\BibitemShut {NoStop}\bibitem [{\citenamefont {Wang}\ \emph {et~al.}(2019)\citenamefont {Wang},
  \citenamefont {Han}, \citenamefont {West}, \citenamefont {Chen},
  \citenamefont {Xie}, \citenamefont {Tian}, \citenamefont {Meunier},
  \citenamefont {Zhang},\ and\ \citenamefont {Li}}]{wang_excitation_2019}  \BibitemOpen
  \bibfield  {author} {\bibinfo {author} {\bibfnamefont {D.}~\bibnamefont
  {Wang}}, \bibinfo {author} {\bibfnamefont {D.}~\bibnamefont {Han}}, \bibinfo
  {author} {\bibfnamefont {D.}~\bibnamefont {West}}, \bibinfo {author}
  {\bibfnamefont {N.-K.}\ \bibnamefont {Chen}}, \bibinfo {author}
  {\bibfnamefont {S.-Y.}\ \bibnamefont {Xie}}, \bibinfo {author} {\bibfnamefont
  {W.~Q.}\ \bibnamefont {Tian}}, \bibinfo {author} {\bibfnamefont
  {V.}~\bibnamefont {Meunier}}, \bibinfo {author} {\bibfnamefont
  {S.}~\bibnamefont {Zhang}}, \ and\ \bibinfo {author} {\bibfnamefont {X.-B.}\
  \bibnamefont {Li}},\ }\href@noop {} {\bibfield  {journal} {\bibinfo
  {journal} {npj Comput. Mater.}\ }\textbf {\bibinfo {volume} {5}},\ \bibinfo
  {pages} {8} (\bibinfo {year} {2019})}\BibitemShut {NoStop}\bibitem [{\citenamefont {Tomasi}\ \emph {et~al.}(2005)\citenamefont {Tomasi},
  \citenamefont {Mennucci},\ and\ \citenamefont {Cammi}}]{PCM-Review}  \BibitemOpen
  \bibfield  {author} {\bibinfo {author} {\bibfnamefont {J.}~\bibnamefont
  {Tomasi}}, \bibinfo {author} {\bibfnamefont {B.}~\bibnamefont {Mennucci}}, \
  and\ \bibinfo {author} {\bibfnamefont {R.}~\bibnamefont {Cammi}},\
  }\href@noop {} {\bibfield  {journal} {\bibinfo  {journal} {Chem. Rev.}\
  }\textbf {\bibinfo {volume} {105}},\ \bibinfo {pages} {2999} (\bibinfo {year}
  {2005})}\BibitemShut {NoStop}\bibitem [{\citenamefont {Marenich}\ \emph {et~al.}(2009)\citenamefont
  {Marenich}, \citenamefont {Cramer},\ and\ \citenamefont {Truhlar}}]{PCM-SMD}  \BibitemOpen
  \bibfield  {author} {\bibinfo {author} {\bibfnamefont {A.~V.}\ \bibnamefont
  {Marenich}}, \bibinfo {author} {\bibfnamefont {C.~J.}\ \bibnamefont
  {Cramer}}, \ and\ \bibinfo {author} {\bibfnamefont {D.~G.}\ \bibnamefont
  {Truhlar}},\ }\href@noop {} {\bibfield  {journal} {\bibinfo  {journal} {J.
  Phys. Chem. B}\ }\textbf {\bibinfo {volume} {113}},\ \bibinfo {pages} {6378}
  (\bibinfo {year} {2009})}\BibitemShut {NoStop}\bibitem [{\citenamefont {Andreussi}\ \emph {et~al.}(2012)\citenamefont
  {Andreussi}, \citenamefont {Dabo},\ and\ \citenamefont {Marzari}}]{PCM-SCCS}  \BibitemOpen
  \bibfield  {author} {\bibinfo {author} {\bibfnamefont {O.}~\bibnamefont
  {Andreussi}}, \bibinfo {author} {\bibfnamefont {I.}~\bibnamefont {Dabo}}, \
  and\ \bibinfo {author} {\bibfnamefont {N.}~\bibnamefont {Marzari}},\
  }\href@noop {} {\bibfield  {journal} {\bibinfo  {journal} {J. Chem. Phys}\
  }\textbf {\bibinfo {volume} {136}},\ \bibinfo {pages} {064102} (\bibinfo
  {year} {2012})}\BibitemShut {NoStop}\bibitem [{\citenamefont {Xiao}\ \emph {et~al.}(2016)\citenamefont {Xiao},
  \citenamefont {Cheng}, \citenamefont {Goddard~III},\ and\ \citenamefont
  {Sundararaman}}]{Goddard}  \BibitemOpen
  \bibfield  {author} {\bibinfo {author} {\bibfnamefont {H.}~\bibnamefont
  {Xiao}}, \bibinfo {author} {\bibfnamefont {T.}~\bibnamefont {Cheng}},
  \bibinfo {author} {\bibfnamefont {W.~A.}\ \bibnamefont {Goddard~III}}, \ and\
  \bibinfo {author} {\bibfnamefont {R.}~\bibnamefont {Sundararaman}},\
  }\href@noop {} {\bibfield  {journal} {\bibinfo  {journal} {J. Am. Chem.
  Soc.}\ }\textbf {\bibinfo {volume} {138}},\ \bibinfo {pages} {483} (\bibinfo
  {year} {2016})}\BibitemShut {NoStop}\bibitem [{\citenamefont {Jason D.~Goodpaster}\ and\ \citenamefont
  {Head-Gordon}(2016)}]{HeadGordon}  \BibitemOpen
  \bibfield  {author} {\bibinfo {author} {\bibfnamefont {A.~T.~B.}\
  \bibnamefont {Jason D.~Goodpaster}}\ and\ \bibinfo {author} {\bibfnamefont
  {M.}~\bibnamefont {Head-Gordon}},\ }\href@noop {} {\bibfield  {journal}
  {\bibinfo  {journal} {J. Phys. Chem. Lett.}\ }\textbf {\bibinfo {volume}
  {7}},\ \bibinfo {pages} {1471} (\bibinfo {year} {2016})}\BibitemShut
  {NoStop}\bibitem [{\citenamefont {Sundararaman}\ \emph {et~al.}(2018)\citenamefont
  {Sundararaman}, \citenamefont {Letchworth-Weaver},\ and\ \citenamefont
  {Schwarz}}]{sundararaman_improving_2018}  \BibitemOpen
  \bibfield  {author} {\bibinfo {author} {\bibfnamefont {R.}~\bibnamefont
  {Sundararaman}}, \bibinfo {author} {\bibfnamefont {K.}~\bibnamefont
  {Letchworth-Weaver}}, \ and\ \bibinfo {author} {\bibfnamefont {K.~A.}\
  \bibnamefont {Schwarz}},\ }\href {\doibase 10.1063/1.5024219} {\bibfield
  {journal} {\bibinfo  {journal} {J. Chem. Phys.}\ }\textbf {\bibinfo {volume}
  {148}},\ \bibinfo {pages} {144105} (\bibinfo {year} {2018})}\BibitemShut
  {NoStop}\bibitem [{\citenamefont {Sundararaman}\ \emph {et~al.}(2014)\citenamefont
  {Sundararaman}, \citenamefont {Gunceler},\ and\ \citenamefont
  {Arias}}]{sundararaman_weighted-density_2014}  \BibitemOpen
  \bibfield  {author} {\bibinfo {author} {\bibfnamefont {R.}~\bibnamefont
  {Sundararaman}}, \bibinfo {author} {\bibfnamefont {D.}~\bibnamefont
  {Gunceler}}, \ and\ \bibinfo {author} {\bibfnamefont {T.~A.}\ \bibnamefont
  {Arias}},\ }\href {\doibase 10.1063/1.4896827} {\bibfield  {journal}
  {\bibinfo  {journal} {J. Chem. Phys.}\ }\textbf {\bibinfo {volume} {141}},\
  \bibinfo {pages} {134105} (\bibinfo {year} {2014})}\BibitemShut {NoStop}\bibitem [{\citenamefont {Gunceler}\ \emph {et~al.}(2013)\citenamefont
  {Gunceler}, \citenamefont {Letchworth-Weaver}, \citenamefont {Sundararaman},
  \citenamefont {Schwarz},\ and\ \citenamefont
  {Arias}}]{gunceler_importance_2013}  \BibitemOpen
  \bibfield  {author} {\bibinfo {author} {\bibfnamefont {D.}~\bibnamefont
  {Gunceler}}, \bibinfo {author} {\bibfnamefont {K.}~\bibnamefont
  {Letchworth-Weaver}}, \bibinfo {author} {\bibfnamefont {R.}~\bibnamefont
  {Sundararaman}}, \bibinfo {author} {\bibfnamefont {K.~A.}\ \bibnamefont
  {Schwarz}}, \ and\ \bibinfo {author} {\bibfnamefont {T.~A.}\ \bibnamefont
  {Arias}},\ }\href {\doibase 10.1088/0965-0393/21/7/074005} {\bibfield
  {journal} {\bibinfo  {journal} {Model. Simul. Mater. Sci. Engi.}\ }\textbf
  {\bibinfo {volume} {21}},\ \bibinfo {pages} {074005} (\bibinfo {year}
  {2013})}\BibitemShut {NoStop}\bibitem [{\citenamefont {Sundararaman}\ \emph {et~al.}(2015)\citenamefont
  {Sundararaman}, \citenamefont {Schwarz}, \citenamefont {Letchworth-Weaver},\
  and\ \citenamefont {Arias}}]{sundararaman_spicing_2015}  \BibitemOpen
  \bibfield  {author} {\bibinfo {author} {\bibfnamefont {R.}~\bibnamefont
  {Sundararaman}}, \bibinfo {author} {\bibfnamefont {K.~A.}\ \bibnamefont
  {Schwarz}}, \bibinfo {author} {\bibfnamefont {K.}~\bibnamefont
  {Letchworth-Weaver}}, \ and\ \bibinfo {author} {\bibfnamefont {T.~A.}\
  \bibnamefont {Arias}},\ }\href {\doibase 10.1063/1.4906828} {\bibfield
  {journal} {\bibinfo  {journal} {J. Chem. Phys.}\ }\textbf {\bibinfo {volume}
  {142}},\ \bibinfo {pages} {054102} (\bibinfo {year} {2015})}\BibitemShut
  {NoStop}\bibitem [{\citenamefont {Sundararaman}\ and\ \citenamefont
  {Goddard}(2015)}]{sundararaman_charge-asymmetric_2015}  \BibitemOpen
  \bibfield  {author} {\bibinfo {author} {\bibfnamefont {R.}~\bibnamefont
  {Sundararaman}}\ and\ \bibinfo {author} {\bibfnamefont {W.~A.}\ \bibnamefont
  {Goddard}},\ }\href {\doibase 10.1063/1.4907731} {\bibfield  {journal}
  {\bibinfo  {journal} {J. Chem. Phys.}\ }\textbf {\bibinfo {volume} {142}},\
  \bibinfo {pages} {064107} (\bibinfo {year} {2015})}\BibitemShut {NoStop}\bibitem [{\citenamefont {Zhang}\ and\ \citenamefont
  {Northrup}(1991)}]{zhang_chemical_1991}  \BibitemOpen
  \bibfield  {author} {\bibinfo {author} {\bibfnamefont {S.}~\bibnamefont
  {Zhang}}\ and\ \bibinfo {author} {\bibfnamefont {J.}~\bibnamefont
  {Northrup}},\ }\href {\doibase 10.1103/PhysRevLett.67.2339} {\bibfield
  {journal} {\bibinfo  {journal} {Phys. Rev. Lett.}\ }\textbf {\bibinfo
  {volume} {67}},\ \bibinfo {pages} {2339} (\bibinfo {year}
  {1991})}\BibitemShut {NoStop}\bibitem [{\citenamefont {Han}\ \emph {et~al.}(2013)\citenamefont {Han},
  \citenamefont {Sun}, \citenamefont {Bang}, \citenamefont {Zhang},
  \citenamefont {Sun}, \citenamefont {Li},\ and\ \citenamefont
  {Zhang}}]{han_deep_2013}  \BibitemOpen
  \bibfield  {author} {\bibinfo {author} {\bibfnamefont {D.}~\bibnamefont
  {Han}}, \bibinfo {author} {\bibfnamefont {Y.~Y.}\ \bibnamefont {Sun}},
  \bibinfo {author} {\bibfnamefont {J.}~\bibnamefont {Bang}}, \bibinfo {author}
  {\bibfnamefont {Y.~Y.}\ \bibnamefont {Zhang}}, \bibinfo {author}
  {\bibfnamefont {H.-B.}\ \bibnamefont {Sun}}, \bibinfo {author} {\bibfnamefont
  {X.-B.}\ \bibnamefont {Li}}, \ and\ \bibinfo {author} {\bibfnamefont {S.~B.}\
  \bibnamefont {Zhang}},\ }\href@noop {} {\bibfield  {journal} {\bibinfo
  {journal} {Phys. Rev. B}\ }\textbf {\bibinfo {volume} {87}},\ \bibinfo
  {pages} {155206} (\bibinfo {year} {2013})}\BibitemShut {NoStop}\bibitem [{\citenamefont {Sundararaman}\ \emph {et~al.}(2017)\citenamefont
  {Sundararaman}, \citenamefont {Letchworth-Weaver}, \citenamefont {Schwarz},
  \citenamefont {Gunceler}, \citenamefont {Ozhabes},\ and\ \citenamefont
  {Arias}}]{sundararaman_jdftx:_2017}  \BibitemOpen
  \bibfield  {author} {\bibinfo {author} {\bibfnamefont {R.}~\bibnamefont
  {Sundararaman}}, \bibinfo {author} {\bibfnamefont {K.}~\bibnamefont
  {Letchworth-Weaver}}, \bibinfo {author} {\bibfnamefont {K.~A.}\ \bibnamefont
  {Schwarz}}, \bibinfo {author} {\bibfnamefont {D.}~\bibnamefont {Gunceler}},
  \bibinfo {author} {\bibfnamefont {Y.}~\bibnamefont {Ozhabes}}, \ and\
  \bibinfo {author} {\bibfnamefont {T.}~\bibnamefont {Arias}},\ }\href
  {\doibase 10.1016/j.softx.2017.10.006} {\bibfield  {journal} {\bibinfo
  {journal} {SoftwareX}\ }\textbf {\bibinfo {volume} {6}},\ \bibinfo {pages}
  {278} (\bibinfo {year} {2017})}\BibitemShut {NoStop}\bibitem [{\citenamefont {Garrity}\ \emph {et~al.}(2014)\citenamefont
  {Garrity}, \citenamefont {Bennett}, \citenamefont {Rabe},\ and\ \citenamefont
  {Vanderbilt}}]{garrity2014pseudopotentials}  \BibitemOpen
  \bibfield  {author} {\bibinfo {author} {\bibfnamefont {K.~F.}\ \bibnamefont
  {Garrity}}, \bibinfo {author} {\bibfnamefont {J.~W.}\ \bibnamefont
  {Bennett}}, \bibinfo {author} {\bibfnamefont {K.~M.}\ \bibnamefont {Rabe}}, \
  and\ \bibinfo {author} {\bibfnamefont {D.}~\bibnamefont {Vanderbilt}},\
  }\href@noop {} {\bibfield  {journal} {\bibinfo  {journal} {Comput. Mater.
  Sci.}\ }\textbf {\bibinfo {volume} {81}},\ \bibinfo {pages} {446} (\bibinfo
  {year} {2014})}\BibitemShut {NoStop}\bibitem [{\citenamefont {Sundararaman}\ and\ \citenamefont
  {Arias}(2013)}]{sundararaman2013regularization}  \BibitemOpen
  \bibfield  {author} {\bibinfo {author} {\bibfnamefont {R.}~\bibnamefont
  {Sundararaman}}\ and\ \bibinfo {author} {\bibfnamefont {T.}~\bibnamefont
  {Arias}},\ }\href@noop {} {\bibfield  {journal} {\bibinfo  {journal} {Phys.
  Rev. B}\ }\textbf {\bibinfo {volume} {87}},\ \bibinfo {pages} {165122}
  (\bibinfo {year} {2013})}\BibitemShut {NoStop}\bibitem [{\citenamefont {Noh}\ \emph {et~al.}(2015)\citenamefont {Noh},
  \citenamefont {Kim}, \citenamefont {Park},\ and\ \citenamefont
  {Kim}}]{noh_deep--shallow_2015}  \BibitemOpen
  \bibfield  {author} {\bibinfo {author} {\bibfnamefont {J.-Y.}\ \bibnamefont
  {Noh}}, \bibinfo {author} {\bibfnamefont {H.}~\bibnamefont {Kim}}, \bibinfo
  {author} {\bibfnamefont {M.}~\bibnamefont {Park}}, \ and\ \bibinfo {author}
  {\bibfnamefont {Y.-S.}\ \bibnamefont {Kim}},\ }\href {\doibase
  10.1103/PhysRevB.92.115431} {\bibfield  {journal} {\bibinfo  {journal} {Phys.
  Rev. B}\ }\textbf {\bibinfo {volume} {92}},\ \bibinfo {pages} {115431}
  (\bibinfo {year} {2015})}\BibitemShut {NoStop}\bibitem [{\citenamefont {Perdew}\ \emph {et~al.}(1996)\citenamefont {Perdew},
  \citenamefont {Burke},\ and\ \citenamefont
  {Ernzerhof}}]{perdew1996generalized}  \BibitemOpen
  \bibfield  {author} {\bibinfo {author} {\bibfnamefont {J.~P.}\ \bibnamefont
  {Perdew}}, \bibinfo {author} {\bibfnamefont {K.}~\bibnamefont {Burke}}, \
  and\ \bibinfo {author} {\bibfnamefont {M.}~\bibnamefont {Ernzerhof}},\
  }\href@noop {} {\bibfield  {journal} {\bibinfo  {journal} {Phys. Rev. Lett.}\
  }\textbf {\bibinfo {volume} {77}},\ \bibinfo {pages} {3865} (\bibinfo {year}
  {1996})}\BibitemShut {NoStop}\bibitem [{\citenamefont {Grimme}(2006)}]{Dispersion-Grimme}  \BibitemOpen
  \bibfield  {author} {\bibinfo {author} {\bibfnamefont {S.}~\bibnamefont
  {Grimme}},\ }\href@noop {} {\bibfield  {journal} {\bibinfo  {journal} {J.
  Comput. Chem}\ }\textbf {\bibinfo {volume} {27}},\ \bibinfo {pages} {1787}
  (\bibinfo {year} {2006})}\BibitemShut {NoStop}\bibitem [{\citenamefont {Freysoldt}\ \emph {et~al.}(2009)\citenamefont
  {Freysoldt}, \citenamefont {Rinke},\ and\ \citenamefont
  {Scheffler}}]{freysoldt2009controlling}  \BibitemOpen
  \bibfield  {author} {\bibinfo {author} {\bibfnamefont {C.}~\bibnamefont
  {Freysoldt}}, \bibinfo {author} {\bibfnamefont {P.}~\bibnamefont {Rinke}}, \
  and\ \bibinfo {author} {\bibfnamefont {M.}~\bibnamefont {Scheffler}},\
  }\href@noop {} {\bibfield  {journal} {\bibinfo  {journal} {Phys. Rev. Lett.}\
  }\textbf {\bibinfo {volume} {103}},\ \bibinfo {pages} {056803} (\bibinfo
  {year} {2009})}\BibitemShut {NoStop}\bibitem [{\citenamefont {H{\"u}ser}\ \emph {et~al.}(2013)\citenamefont
  {H{\"u}ser}, \citenamefont {Olsen},\ and\ \citenamefont
  {Thygesen}}]{huser2013quasiparticle}  \BibitemOpen
  \bibfield  {author} {\bibinfo {author} {\bibfnamefont {F.}~\bibnamefont
  {H{\"u}ser}}, \bibinfo {author} {\bibfnamefont {T.}~\bibnamefont {Olsen}}, \
  and\ \bibinfo {author} {\bibfnamefont {K.~S.}\ \bibnamefont {Thygesen}},\
  }\href@noop {} {\bibfield  {journal} {\bibinfo  {journal} {Phys. Rev. B}\
  }\textbf {\bibinfo {volume} {87}},\ \bibinfo {pages} {235132} (\bibinfo
  {year} {2013})}\BibitemShut {NoStop}\bibitem [{\citenamefont {Abdi}\ \emph {et~al.}(2018)\citenamefont {Abdi},
  \citenamefont {Chou}, \citenamefont {Gali},\ and\ \citenamefont
  {Plenio}}]{abdi2018color}  \BibitemOpen
  \bibfield  {author} {\bibinfo {author} {\bibfnamefont {M.}~\bibnamefont
  {Abdi}}, \bibinfo {author} {\bibfnamefont {J.-P.}\ \bibnamefont {Chou}},
  \bibinfo {author} {\bibfnamefont {A.}~\bibnamefont {Gali}}, \ and\ \bibinfo
  {author} {\bibfnamefont {M.~B.}\ \bibnamefont {Plenio}},\ }\href@noop {}
  {\bibfield  {journal} {\bibinfo  {journal} {ACS Photonics}\ }\textbf
  {\bibinfo {volume} {5}},\ \bibinfo {pages} {1967} (\bibinfo {year}
  {2018})}\BibitemShut {NoStop}\bibitem [{\citenamefont {Reimers}\ \emph {et~al.}(2018)\citenamefont
  {Reimers}, \citenamefont {Sajid}, \citenamefont {Kobayashi},\ and\
  \citenamefont {Ford}}]{reimers2018understanding}  \BibitemOpen
  \bibfield  {author} {\bibinfo {author} {\bibfnamefont {J.~R.}\ \bibnamefont
  {Reimers}}, \bibinfo {author} {\bibfnamefont {A.}~\bibnamefont {Sajid}},
  \bibinfo {author} {\bibfnamefont {R.}~\bibnamefont {Kobayashi}}, \ and\
  \bibinfo {author} {\bibfnamefont {M.~J.}\ \bibnamefont {Ford}},\ }\href@noop
  {} {\bibfield  {journal} {\bibinfo  {journal} {J. Chem. Theory Comput.}\
  }\textbf {\bibinfo {volume} {14}},\ \bibinfo {pages} {1602} (\bibinfo {year}
  {2018})}\BibitemShut {NoStop}\bibitem [{\citenamefont {Popescu}\ and\ \citenamefont
  {Zunger}(2012)}]{popescu2012extracting}  \BibitemOpen
  \bibfield  {author} {\bibinfo {author} {\bibfnamefont {V.}~\bibnamefont
  {Popescu}}\ and\ \bibinfo {author} {\bibfnamefont {A.}~\bibnamefont
  {Zunger}},\ }\href@noop {} {\bibfield  {journal} {\bibinfo  {journal} {Phys.
  Rev. B}\ }\textbf {\bibinfo {volume} {85}},\ \bibinfo {pages} {085201}
  (\bibinfo {year} {2012})}\BibitemShut {NoStop}\end{thebibliography}
\makeatletter{}

\end{document}